\documentclass[twocolumn,aps,prc,superscriptaddress]{revtex4-1}

\usepackage{color,amsmath,amssymb,epsfig,graphicx}
\usepackage{bm}
\usepackage{mathptmx, courier, pifont}
\usepackage[scaled=0.92]{helvet}
\usepackage[T1]{fontenc}
\usepackage{textcomp}
\usepackage[colorlinks=true,linkcolor=blue,filecolor=blue,urlcolor=blue,citecolor=blue]{hyperref}

\def\bra{\,<\!} \def\ket{\!>\,} \def\ack{\,|\,}

\begin{document}
\author{S. Jehangir}
\email{sheikhahmad.phy@gmail.com}
\affiliation{Department of Physics, Islamic University of Science and Technology,
 Jammu and Kashmir, 192 122, India}
\author{Nazira Nazir}
\affiliation{Department of Physics, University of Kashmir,
  Hazratbal, Srinagar, 190 006, India}
\author{G. H. Bhat}
\email{gwhr.bhat@gmail.com}
\affiliation{Department of Physics, S.P. College,  Srinagar, Jammu and Kashmir, 190 001, India}
\affiliation{Cluster University Srinagar, Jammu and Kashmir,
  Srinagar, Goji Bagh, 190 008, India}
\author{J. A. Sheikh}
\email{sjaphysics@gmail.com}
\affiliation{Department of Physics, University of Kashmir,
  Hazratbal, Srinagar, 190 006, India}
\author{N. Rather}
\affiliation{Department of Physics, Islamic University of Science and Technology, 
Jammu and Kashmir, 192 122, India}
\author{S. Chakraborty}
\affiliation{Physics Group, Variable Energy Cyclotron Centre, 
Kolkata, 700 064, India}
\author{R. Palit}
\affiliation{Department of Nuclear and Atomic Physics, 
  Tata Institute of Fundamental Research, Mumbai, 400 005, India}
\title{Extended triaxial projected shell model approach for
  odd-neutron nuclei}

\begin{abstract}

In an effort to elucidate the rich band structures observed in odd-neutron systems, triaxial
projected shell model approach is extended to
include three-quasineutron and five-quasiparticle configurations. This extension
makes it possible to investigate the high-spin states up to and including the second band crossing.
Detailed investigation has been performed for odd-mass Xe isotopes
with the extended basis, and it is shown that character of the band crossing along the yrast line changes with
shell filling of the $1h_{11/2}$ orbital. Further, it is observed that the three-quasiparticle
state that crosses the ground-state configuration,
leading to the normal band crossing phenomenon along the yrast line, first crosses the
$\gamma$ band based on the ground-state configuration at an earlier spin value. This
crossing feature explains the occurrence of the
signature inversion observed in the $\gamma$ bands for some of the studied isotopes.
\end{abstract}

\date{\today}

\maketitle

\section{\label{sec:Intro}Introduction}
In recent years, some major advancements in the spectroscopic
techniques has made it feasible to populate the high-spin band structures 
in atomic nuclei to the extremes of angular-momentum, excitation
energy and isospin \cite{gammasphere,AGATA,EUROBALL,
120Te,122Te,123I,125I,123Xe,124Xe,125Xe,126Xe,119Cs,156Dy,167Lu,168Yb,197_198Hg,195Hg,198Po}. 
In some nuclei, the high-spin states have been studied up to angular momentum, 
$ I \approx 60\hbar $ and as many as twenty side
bands have been identified \cite{124Xe,125Xe,126Xe,119Cs,167Lu}. The 
observation of these rich
band structures poses a tremendous challenge to theoretical models to 
elucidate the properties of these structures. During the last several
decades, the standard approach to describe the high-spin
properties of deformed nuclei has been the cranked shell model (CSM) 
based on modified harmonic oscillator \cite{Ben79aw, Ben79w} or Woods-Saxon potential \cite{PhysRevC.23.920}.
Although this approach has provided some new insights into the structure
of the high-spin states, but is known to have limited
applicability. For instance, it is suited only for rotating
systems, and the study of vibrational modes is beyond the scope of the
basic CSM approach. Further, the CSM wave-functions don't 
have well defined value of the angular-momentum, and the evaluation of the
transition probabilities using this approach becomes questionable \cite{RS80}.  

The spherical shell model (SSM) approach has made great strides in
recent years to  explore the medium and heavier mass regions, and it
has  become possible to investigate the properties of nuclei in the
mass region, $ A \approx 120-130 $ \cite{RevModPhys.92.015002,63-67Ga,60-66Zn,63Cu,135La,66Zn,132Te,119-126Sn,116Sn}. 
However, in order to study the high-spin band structures, it is essential 
to include, at-least, the configuration space of
two oscillator shells, which seems to be impossible with the
existing computational facilities. The modern mean-field 
approaches based on Skyrme, Gogny and relativistic  density
functional, on the other hand, reproduce the known binding energies
of nuclei all across the Segr\`e-chart with a remarkable accuracy
\cite{je12,NS19,FR20,SG09,(Rob19),(Rin96b)}.  The problem with these approaches is that they are mostly
limited to investigate the ground-state properties as beyond the
mean-field extensions using the projection methods leads to the
singularities \cite{PhysRevC.76.054315,PhysRevC.79.044320,JAS21,MB03}, and further one encounters conceptual
problem on how to treat the density-dependent terms in
the functionals of these approaches. In recent years, alternative
approaches have been developed to map the energy surfaces obtained
from the density functional onto Bohr \cite{Alh06,Alh08,PhysRevC.84.014302} and interacting boson
model (IBM) \cite{PhysRevC.104.024323,PhysRevC.104.054320} Hamiltonians, and fit the parameters of these
phenomenological approaches. These model Hamiltonians are then solved
using the standard techniques to obtain the 
energies and electromagnetic properties of the nuclear systems. However, these alternative approaches are restricted, at the
moment, to investigate ground-state band structures only.

\begin{figure*}[htb]
\vspace{1cm}
\includegraphics[totalheight=14.5cm]{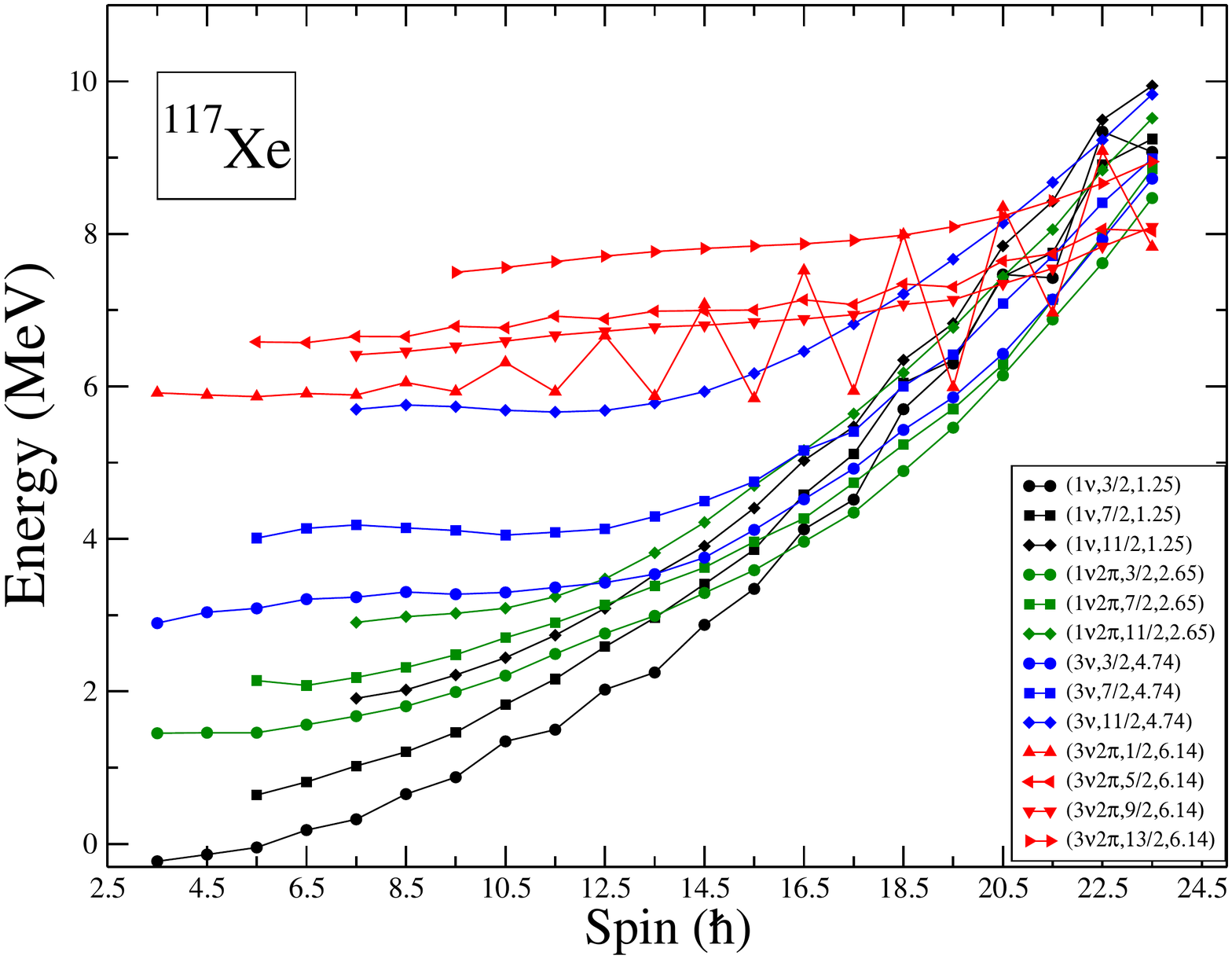} 
\caption{
Angular-momentum projected energies are shown before diagonalization of the
shell model Hamiltonian for $^{117}$Xe. The bands are labelled by
three quantities : group structure, $K$-quantum number and energy of the quasiparticle
state. For instance, $(1\nu,3/2,1.25)$ designates one-quasineutron
state having intrinsic energy of 1.25 MeV and $K=3/2$. It is interesting to note that apart from the
normal band crossing at $I=33/2$ between the three-quasiparticle configuration having energy of 2.65 MeV and $K=3/2$ with
the ground state band, the $\gamma$ band built on the three-quasiparticle state having $K=7/2$ also
crosses the $\gamma$ band based on the ground state at the same angular momentum. Further, the three-quasiparticle
configuration first crosses the $\gamma$ band around $I=14$ before it crosses the ground state band. This crossing
leads to signature inversion phenomenon in the $\gamma$ band.}
\label{bd1}
\end{figure*}
\begin{figure}[htb]
 \centerline{\includegraphics[totalheight=13cm]{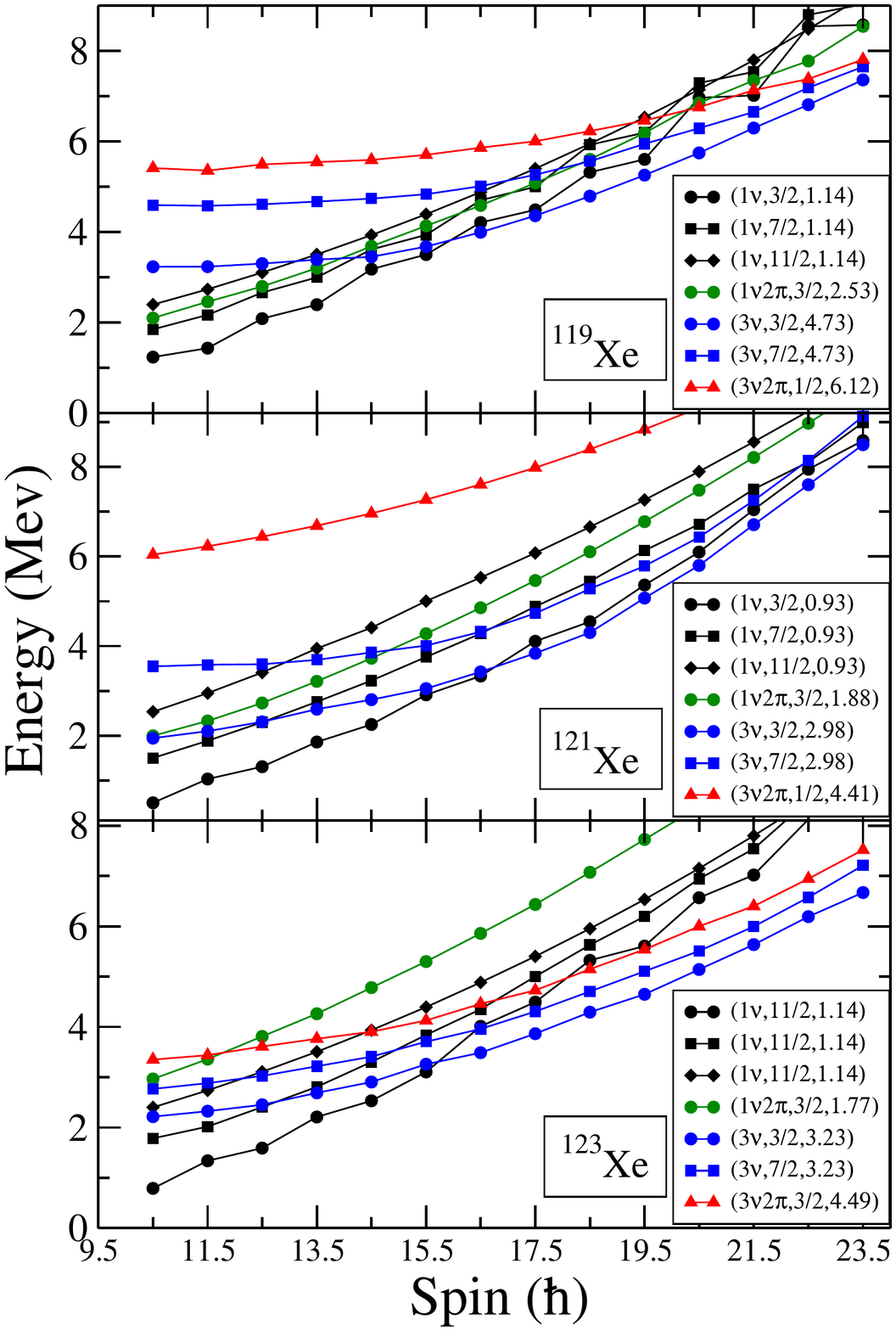}}
\caption{
Segments of the band diagrams for odd-neutron $^{119-123}$Xe isotopes. The
bands are labelled as in Fig.~\ref{bd1}.} \label{bd2a}
\end{figure}
\begin{figure}[htb]
 \centerline{\includegraphics[trim=0cm 0cm 0cm
0cm,width=0.55\textwidth,clip]{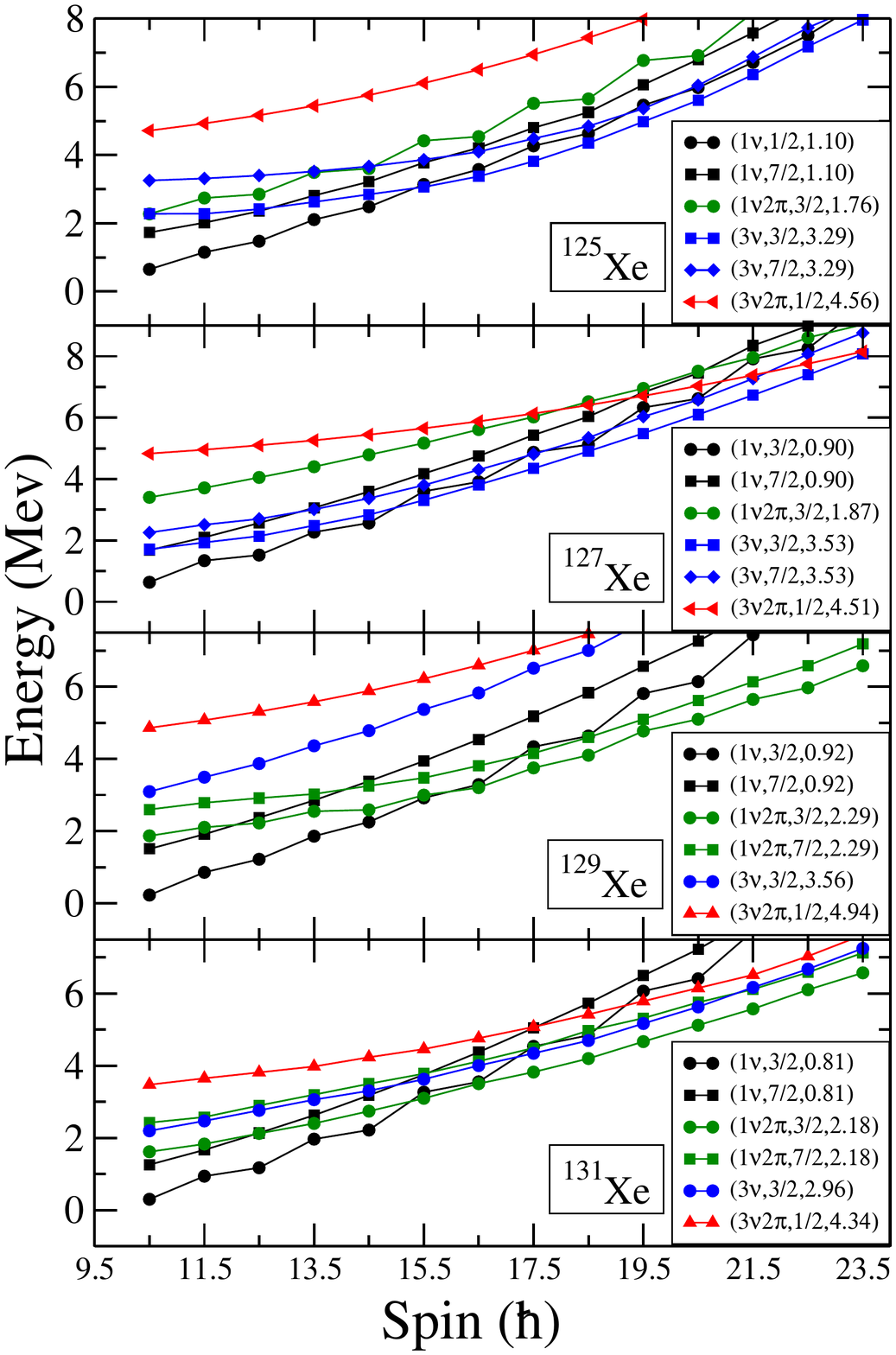}}
\caption{
Segments of the band diagrams for odd-neutron $^{125-131}$Xe isotopes. The
bands are labelled as in Fig.~\ref{bd1}.} \label{bd2b}
\end{figure}



\begin{figure*}[ht]
\includegraphics[width=\textwidth]{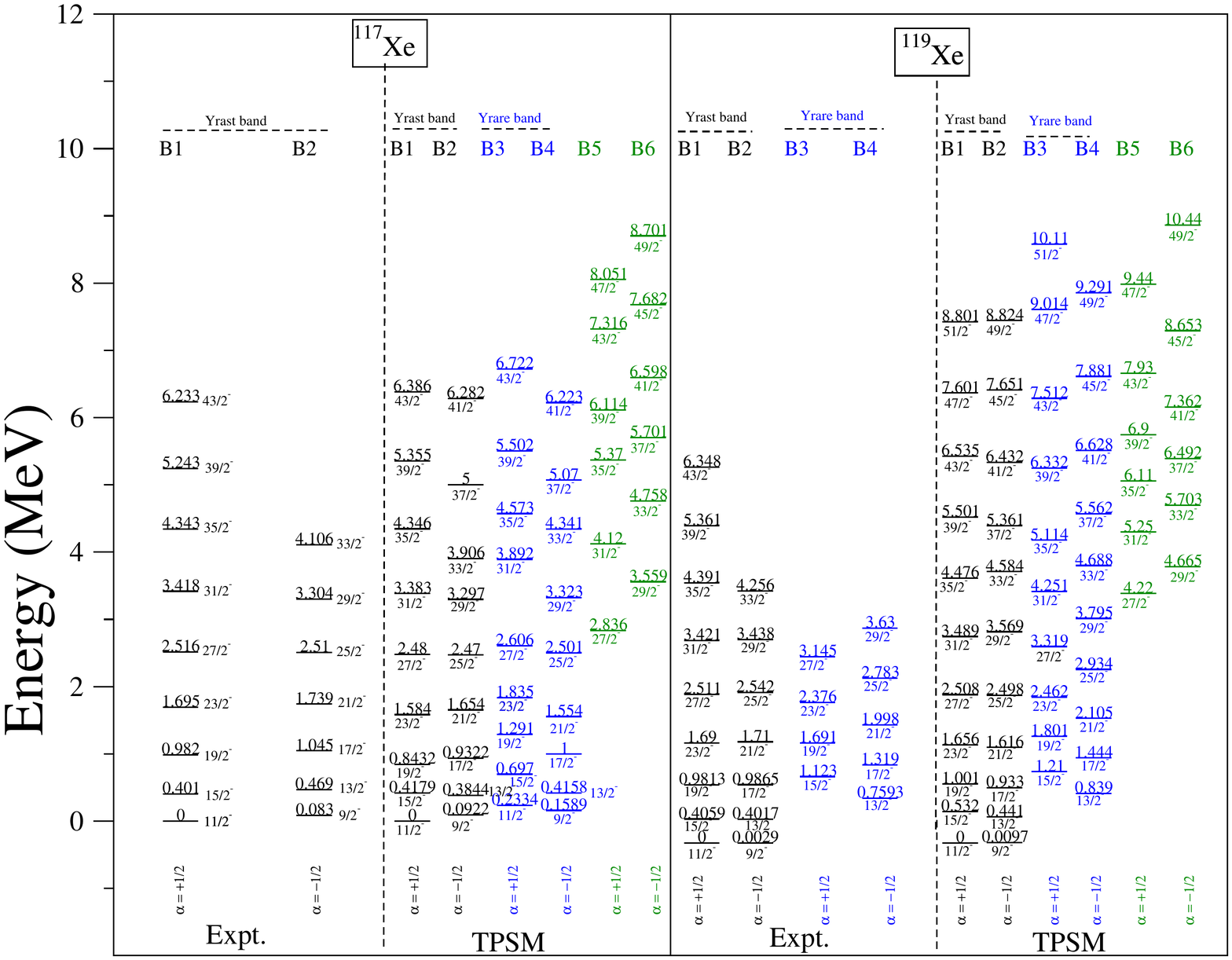}
\caption{TPSM energies for the lowest bands after configuration
mixing are plotted along
with the available experimental data for  $^{117,119}$Xe  isotopes. Data is
taken from \cite{117xeb,119xeb}.}
\label{expe1}
\end{figure*}
\begin{figure*}[htb]
\vspace{0cm}
 {\includegraphics[totalheight=13cm]{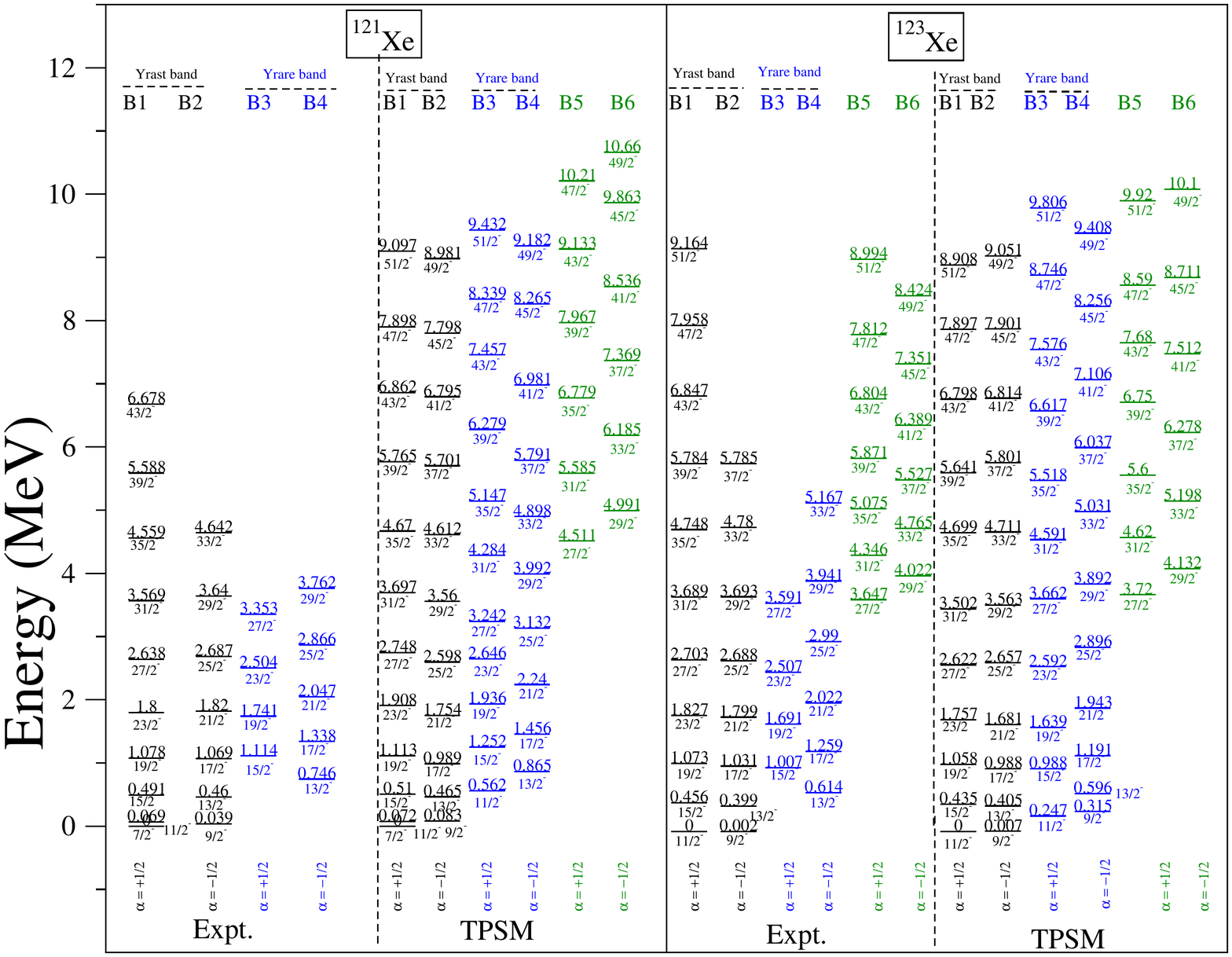}} 
\caption{
TPSM energies for the lowest  bands after configuration
mixing are plotted along
with the available experimental data for  $^{121,123}$Xe  isotopes. Data  is
taken from \cite{Tim_r_1995,123Xe}. } \label{expe2}
\end{figure*}
\begin{figure*}[htb]
\vspace{0cm}
\includegraphics[totalheight=13cm]{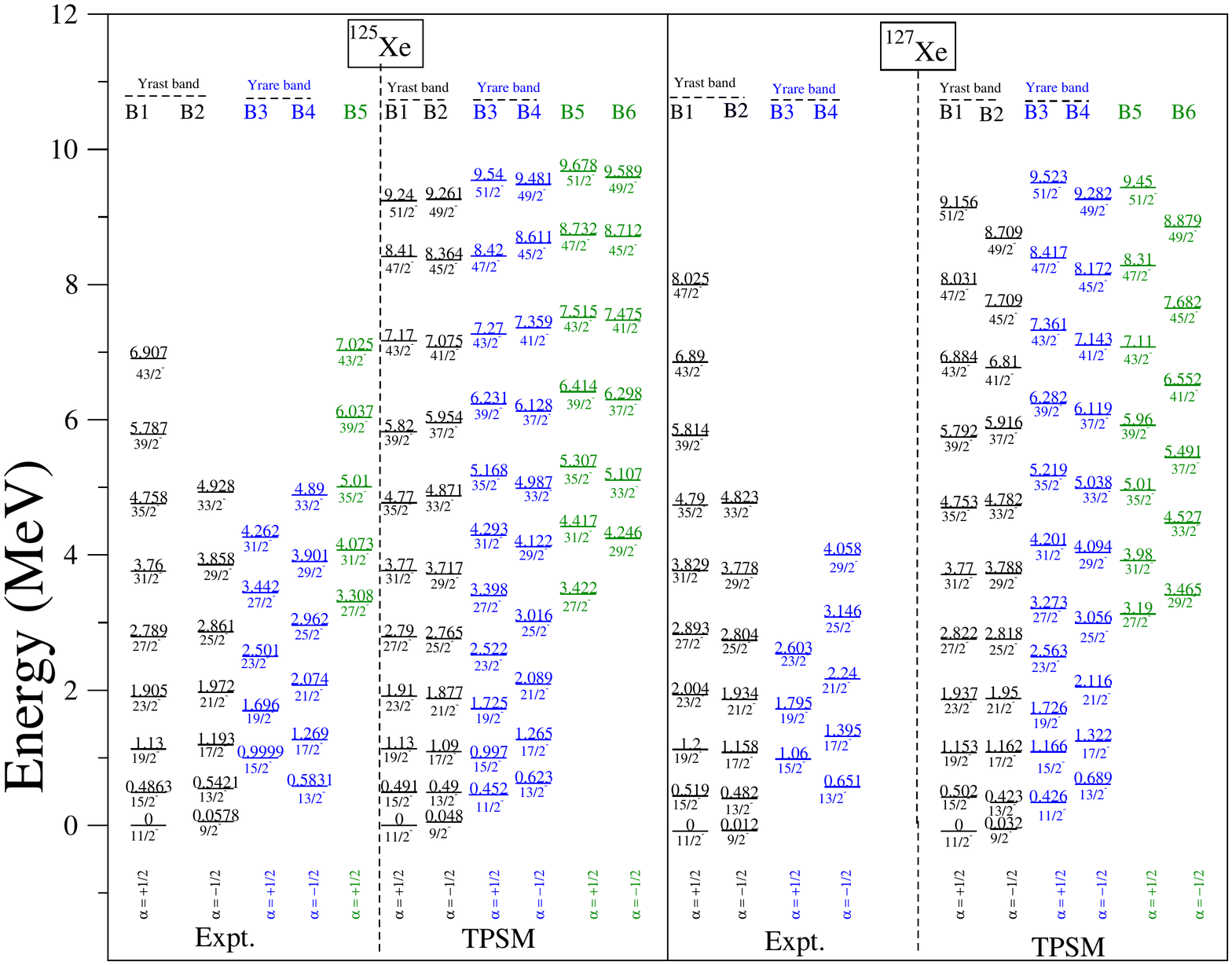} 
\caption{
TPSM energies for the lowest  bands after configuration
mixing are plotted along
with the available experimental data for  $^{125,127}$Xe  isotopes. Data is
taken from\cite{125Xe,127Xe-2020,127Xech}.} \label{expe3}
\end{figure*}
\begin{figure*}[htb]
\vspace{0cm}
\includegraphics[totalheight=13cm]{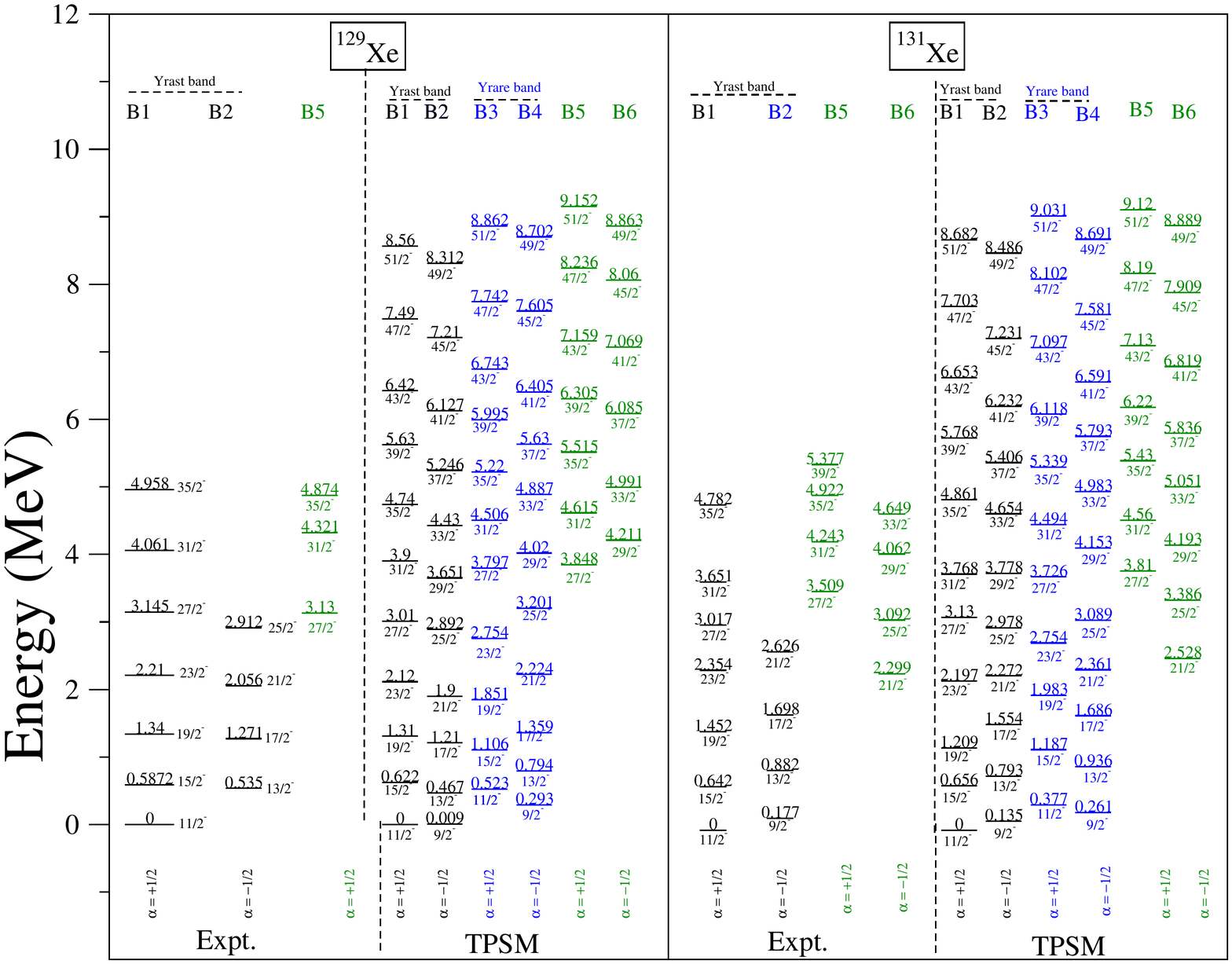} 
\caption{
TPSM energies for the lowest  bands after configuration
mixing are plotted along
with the available experimental data for  $^{129,131}$Xe  isotopes. Data  is
taken from \cite{129xed,131xed,131prxe}.} \label{expe4}
\end{figure*}
\begin{figure}[htb]
 \centerline{\includegraphics[trim=0cm 0cm 0cm
0cm,width=0.53\textwidth,clip]{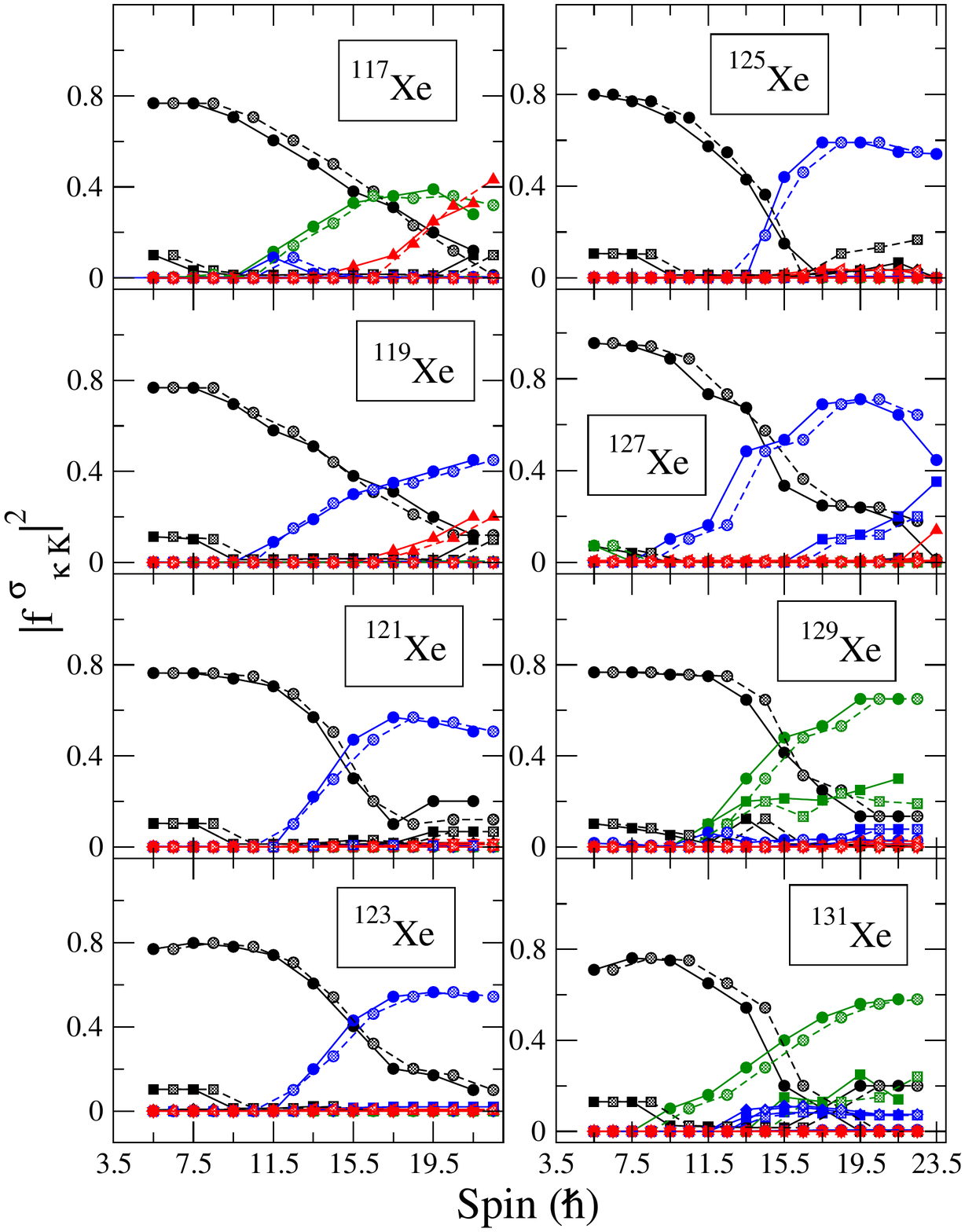}}
\caption{
Amplitudes of various projected   $K$-configurations in the wave
functions of the yrast band after diagonalization for $^{117-131}$Xe isotopes. 
The curves are labelled as in Fig.~\ref{bd1}. It needs to be clarified
that the projected basis states are not orthogonal, and the amplitudes displayed are not
probabilities in the true sense.} \label{wf2}
\end{figure}

\begin{figure}[htb]
\centerline{\includegraphics[trim=0cm 0cm 0cm
0cm,width=0.53\textwidth,clip]{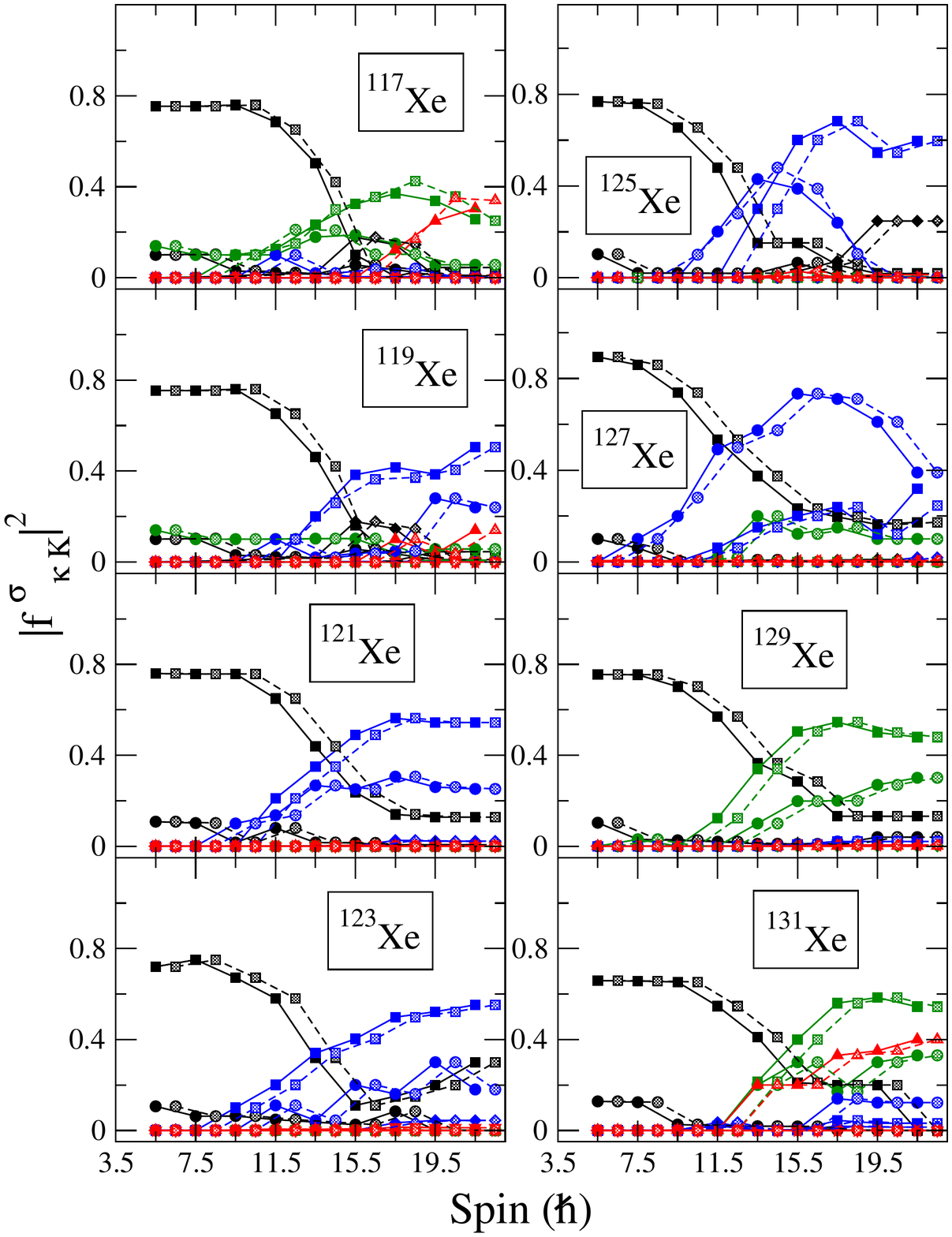}}
\caption{
Amplitudes of various projected $K$-configurations in the wave
functions of the yrare band after diagonalization for $^{117-131}$Xe isotopes. 
The curves are labelled as in Fig.~\ref{bd1}.} \label{wf3}
\end{figure}

\begin{figure}[htb]
 \centerline{\includegraphics[trim=0cm 0cm 0cm
0cm,width=0.53\textwidth,clip]{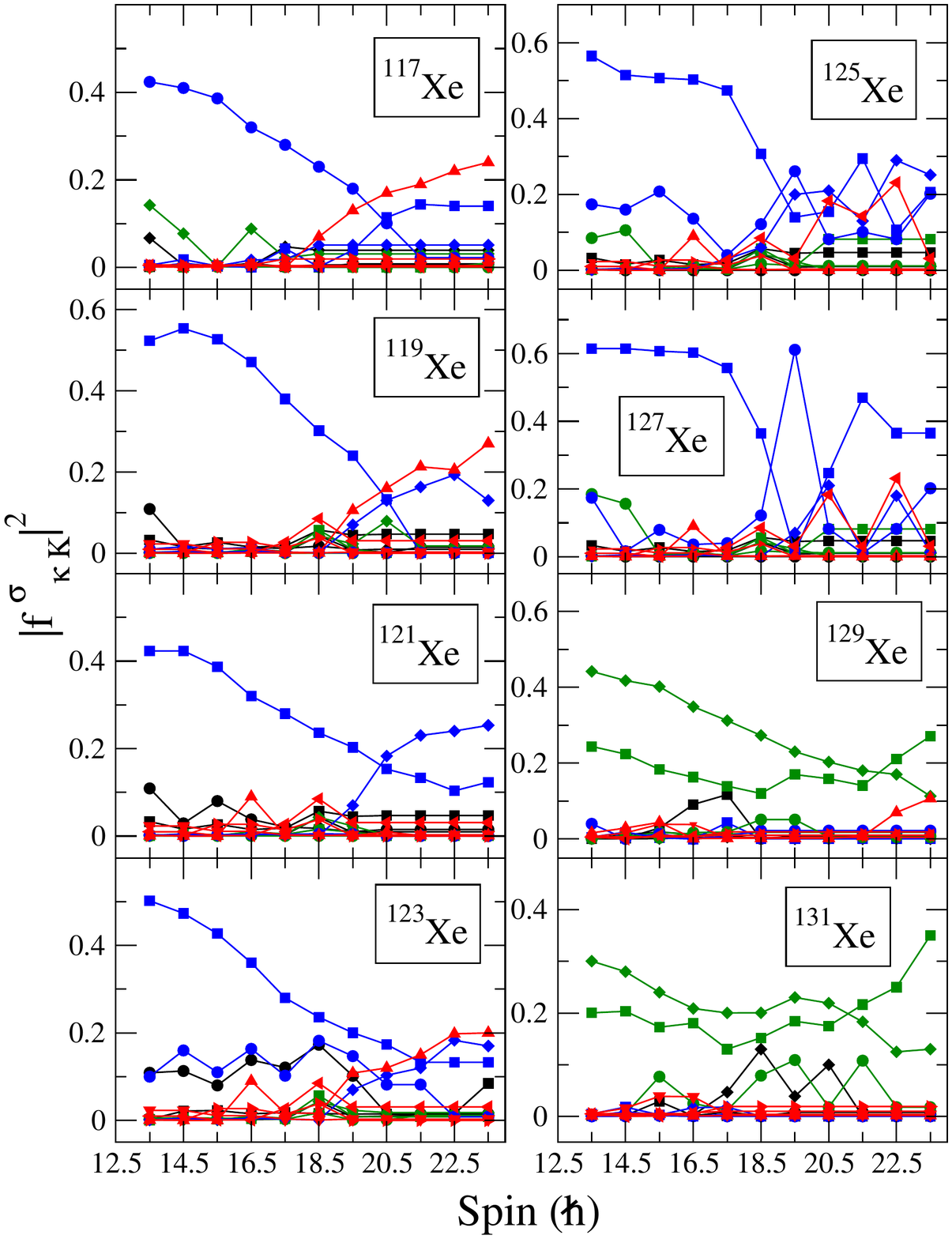}}
\caption{
Amplitudes of various projected $K$-configurations in the wave functions of the second excited band 
after diagonalization for $^{117-131}$Xe isotopes. The curves are labelled as in Fig.~\ref{bd1}.} 
\label{wf4}
\end{figure}

\begin{figure}[htb]
 \centerline{\includegraphics[trim=0cm 0cm 0cm
0cm,width=0.53\textwidth,clip]{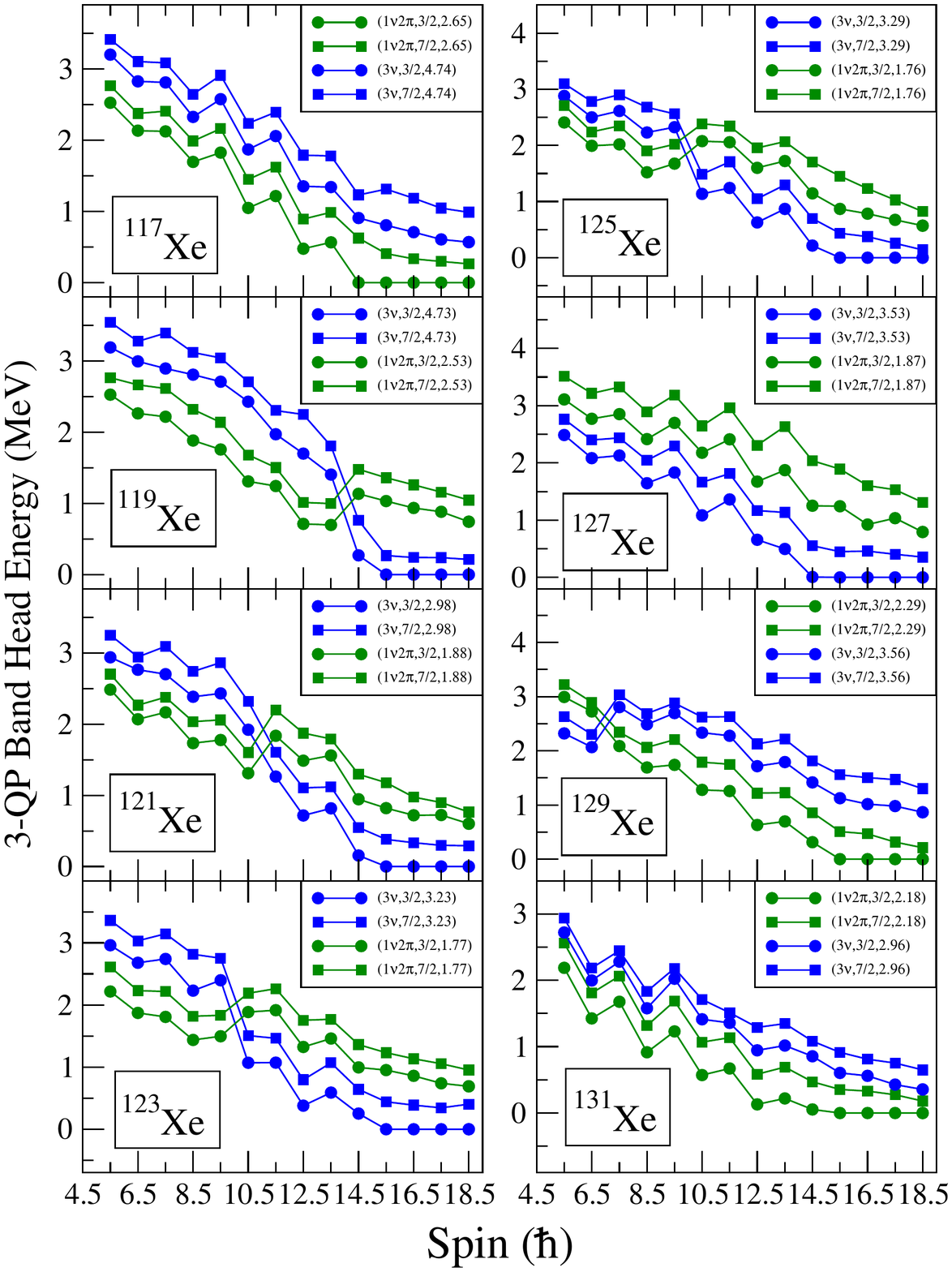}}
\caption{
Lowest few calculated 3-qp band head energies after configuration mixing for
odd-neutron $^{117-131}$Xe  isotopes. The band head energies are plotted with respect to the
lowest state at each angular momentum.} \label{bhe1}
\end{figure}

\begin{figure}[htb]
 \centerline{\includegraphics[trim=0cm 0cm 0cm
0cm,width=0.53\textwidth,clip]{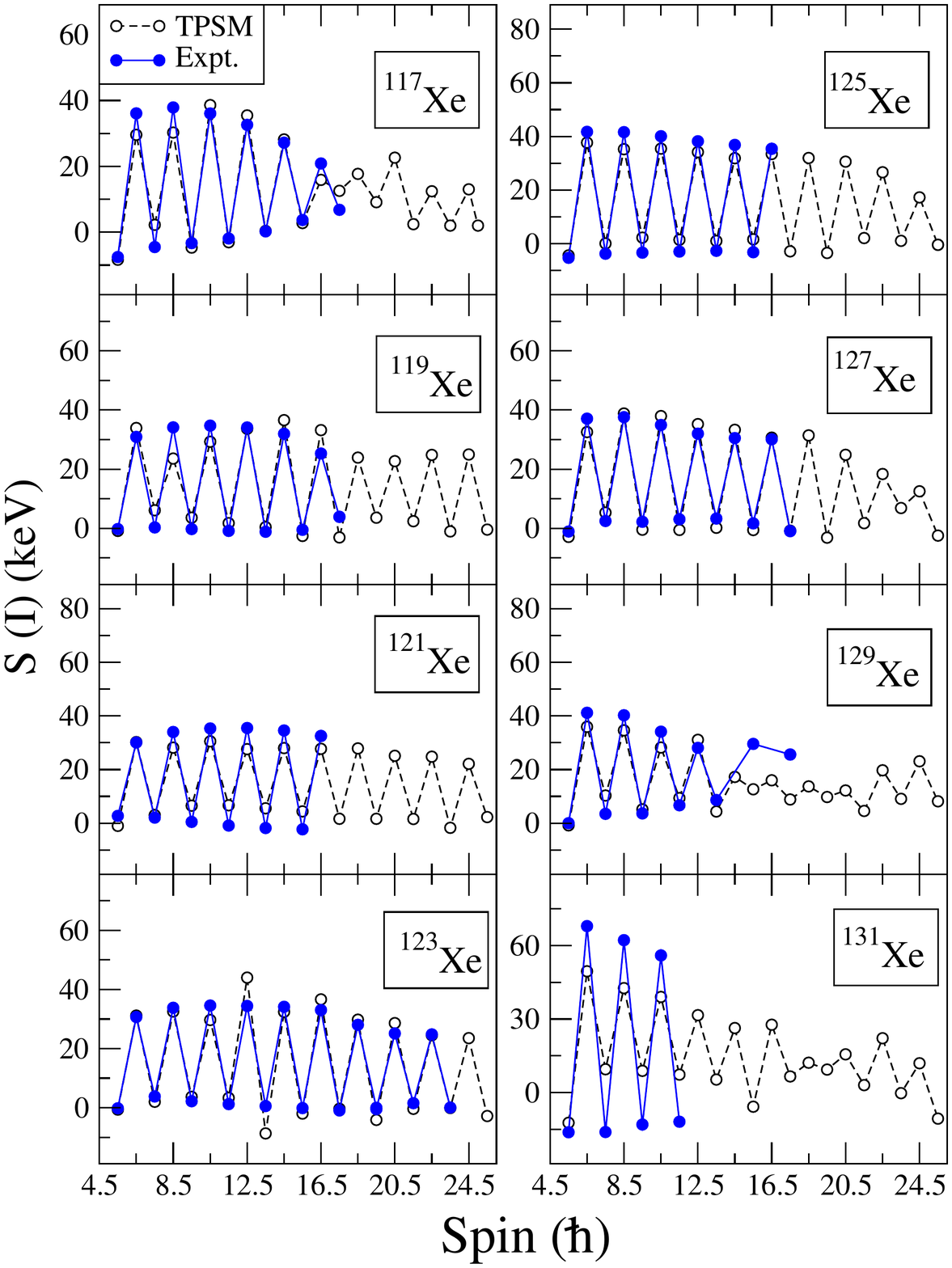}}
\caption{
Plots of the signature splitting of the negative-parity yrast-bands in 
$^{117-131}$Xe nuclei with $ S(I) = (E(I)-E(I-1))/2I $.} \label{sp1y}
\end{figure}
\begin{figure}[htb]
 \centerline{\includegraphics[trim=0cm 0cm 0cm
0cm,width=0.53\textwidth,clip]{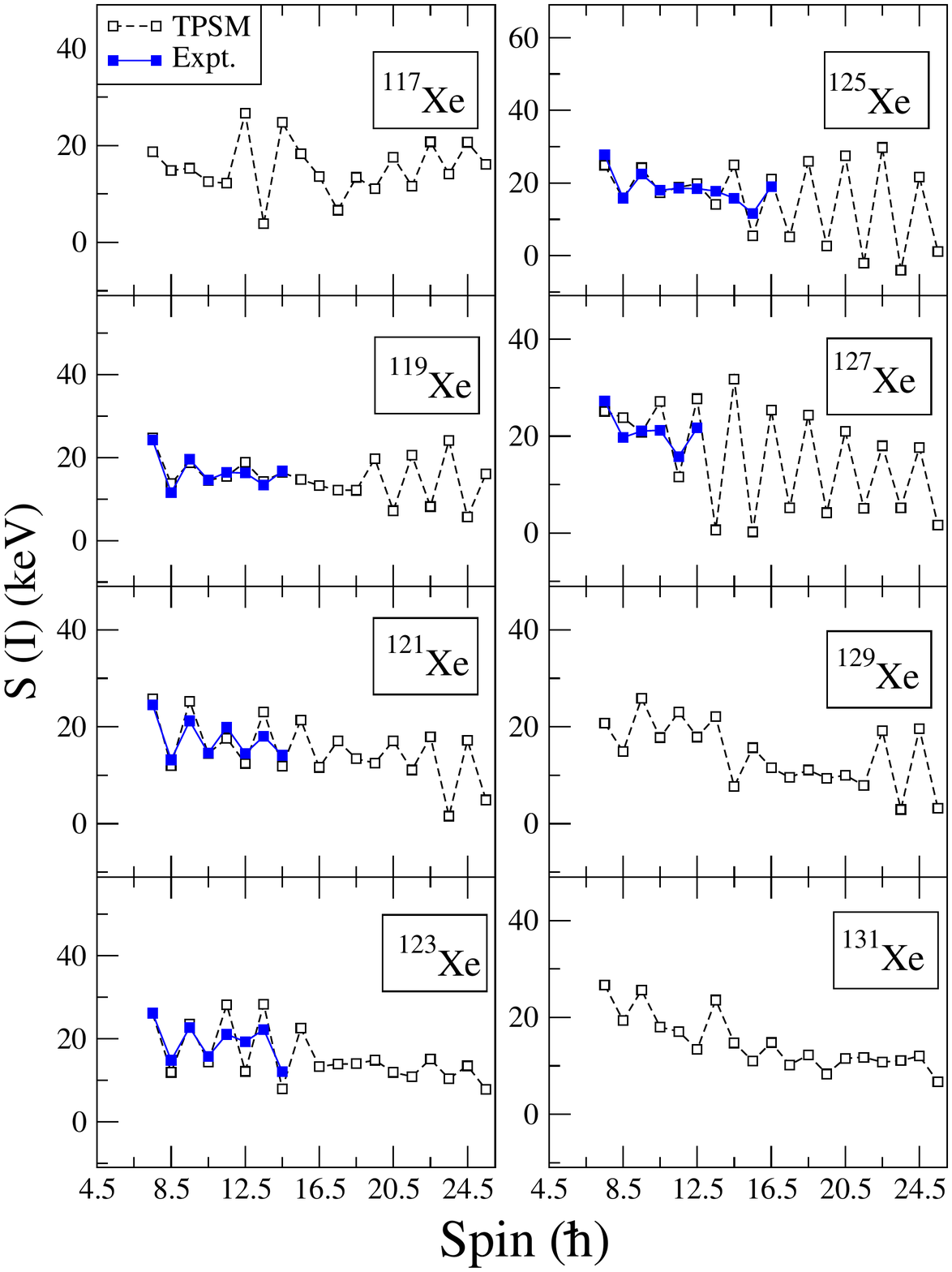}}
\caption{
Plots of the signature splitting of the negative-parity yrare-bands in 
$^{117-131}$Xe nuclei with $ S(I) = (E(I)-E(I-1))/2I $.} \label{sp2y}
\end{figure}

\begin{figure}[htb]
 \centerline{\includegraphics[trim=0cm 0cm 0cm
0cm,width=0.53\textwidth,clip]{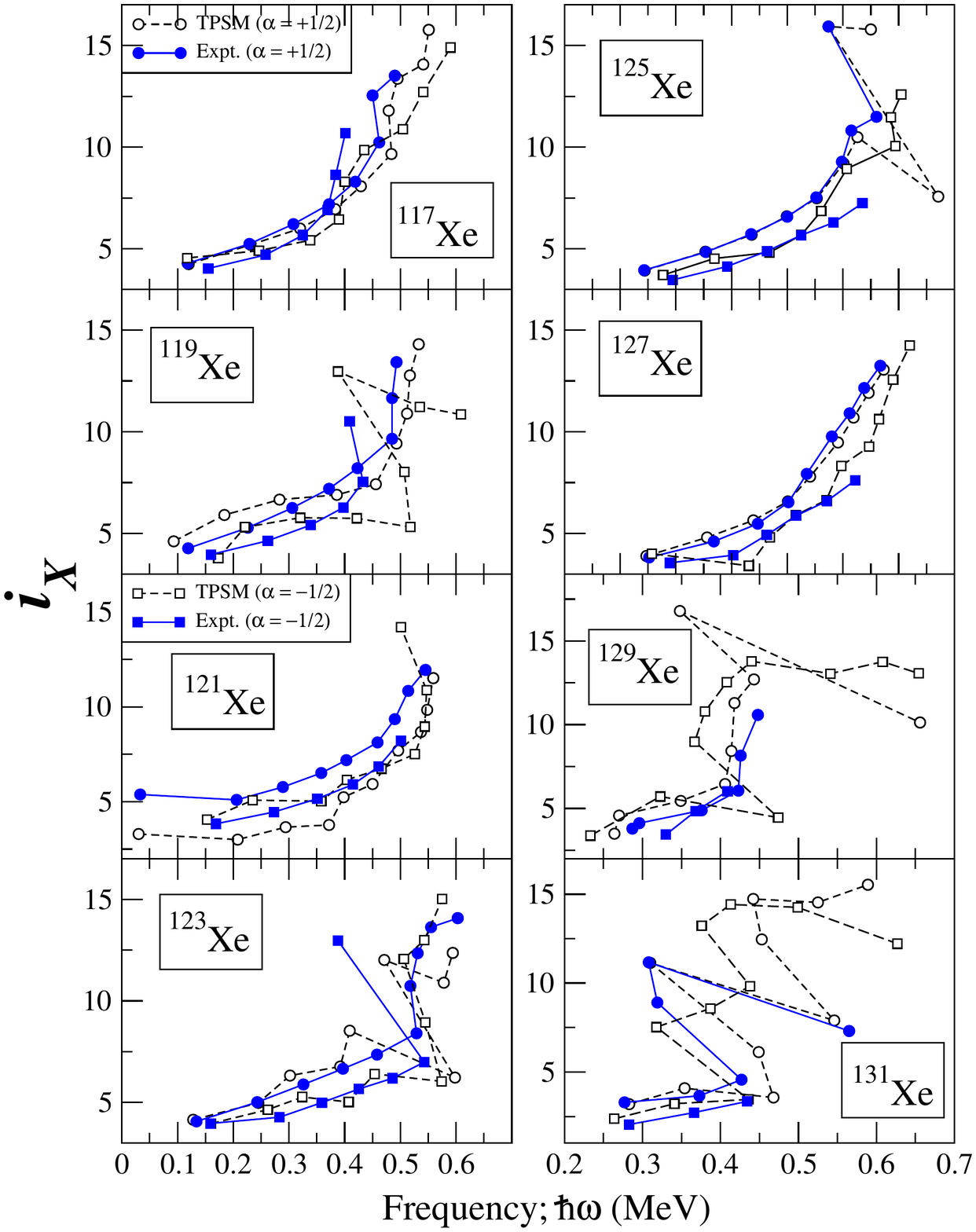}}
\caption{
Comparison of the aligned angular momentum,  $i_x=I_x(\omega)-I_{x,ref}(\omega)$, 
where $\hbar\omega=\frac{E_{\gamma}}{I_x^i(\omega)-I_x^f(\omega)}$,  
$I_x(\omega)= \sqrt{I(I+1)-K^2}$  and $I_{x,ref}(\omega)=\omega(J_0+\omega^{2}J_1)$. 
The reference band Harris parameters used are $J_0$=23 and $J_1$=90, obtained from the measured 
energy levels as well as those calculated from the TPSM results, for $^{117-131}$Xe nuclei.} 
\label{ali1}
\end{figure}
\begin{figure}[htb]
 \centerline{\includegraphics[trim=0cm 0cm 0cm
0cm,width=0.53\textwidth,clip]{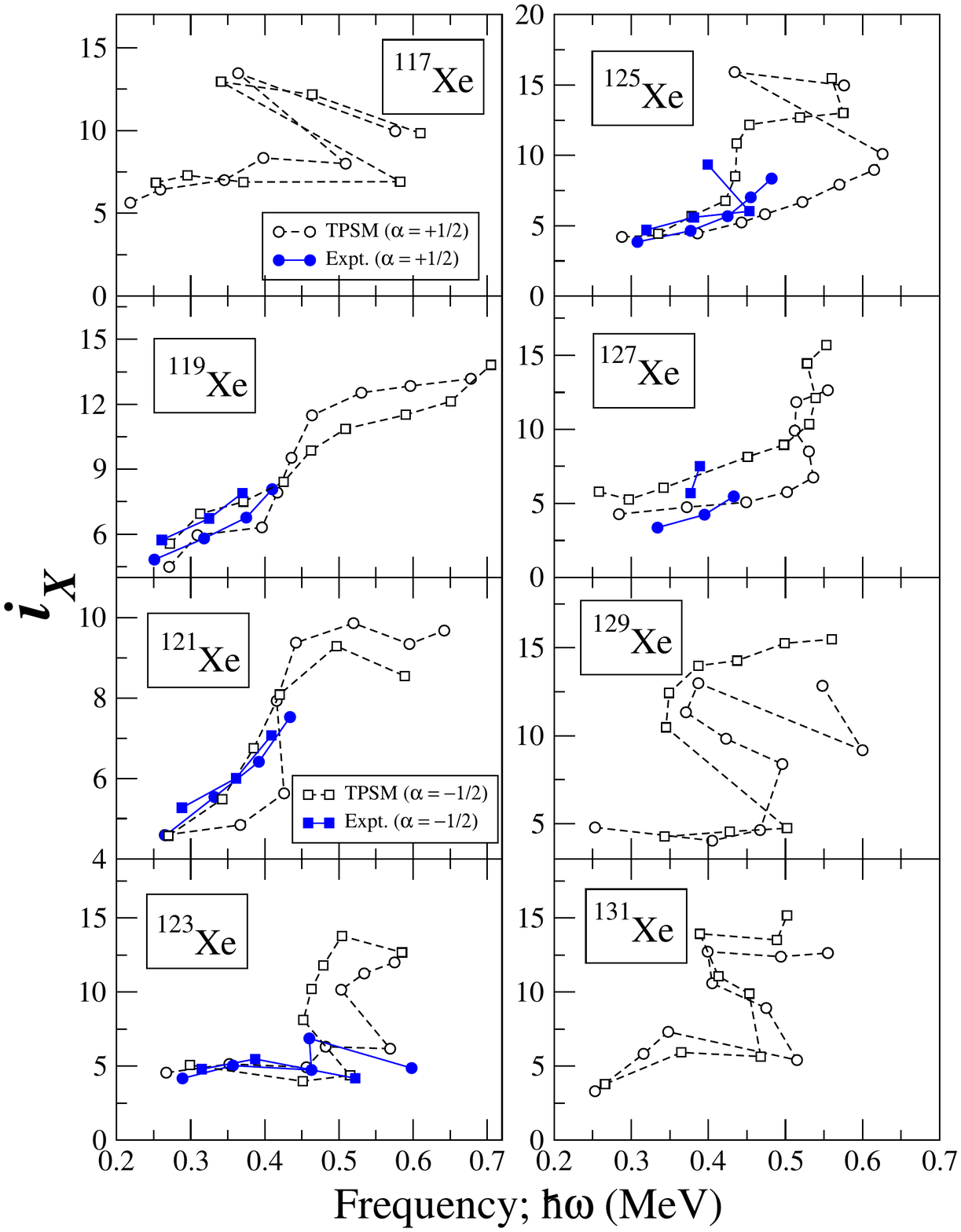}}
\caption{
Comparison of the aligned angular momentum for yrare-band obtained from the measured 
energy levels as well as those calculated from the TPSM results, for $^{117-131}$Xe nuclei.} 
\label{ali1yrare}
\end{figure}
\begin{figure}[htb]
 \centerline{\includegraphics[trim=0cm 0cm 0cm
0cm,width=0.53\textwidth,clip]{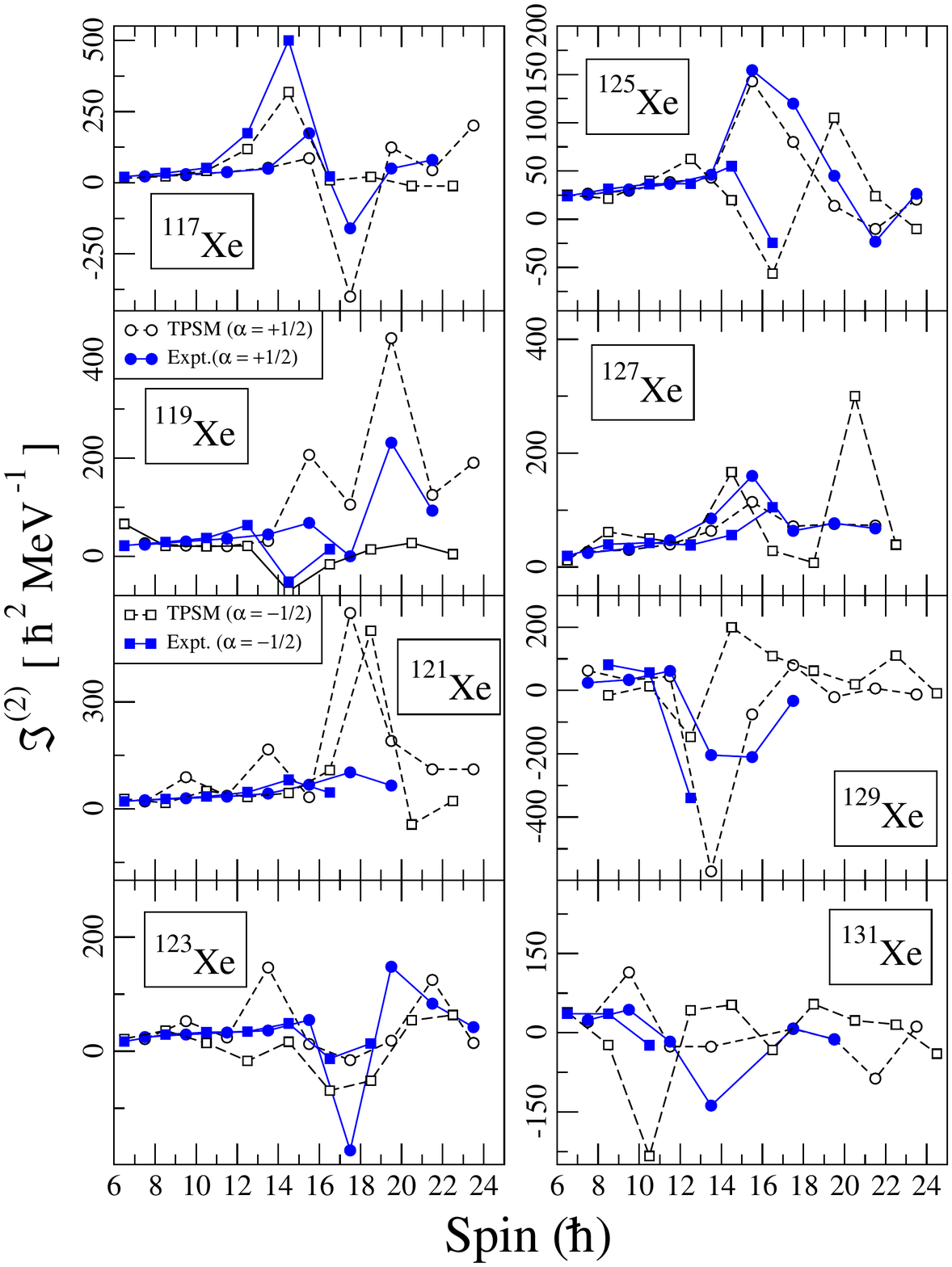}}
\caption{
Comparison between experimental and calculated dynamic moment of inertia,  
$J^{(2)} =  \frac{4}{E_{\gamma}(I)-E_{\gamma}(I-2)}$, of the yrast-band for$^{117-131}$Xe isotopes. }
\label{ali2}
\end{figure}
\begin{figure}[htb]
 \centerline{\includegraphics[trim=0cm 0cm 0cm
0cm,width=0.53\textwidth,clip]{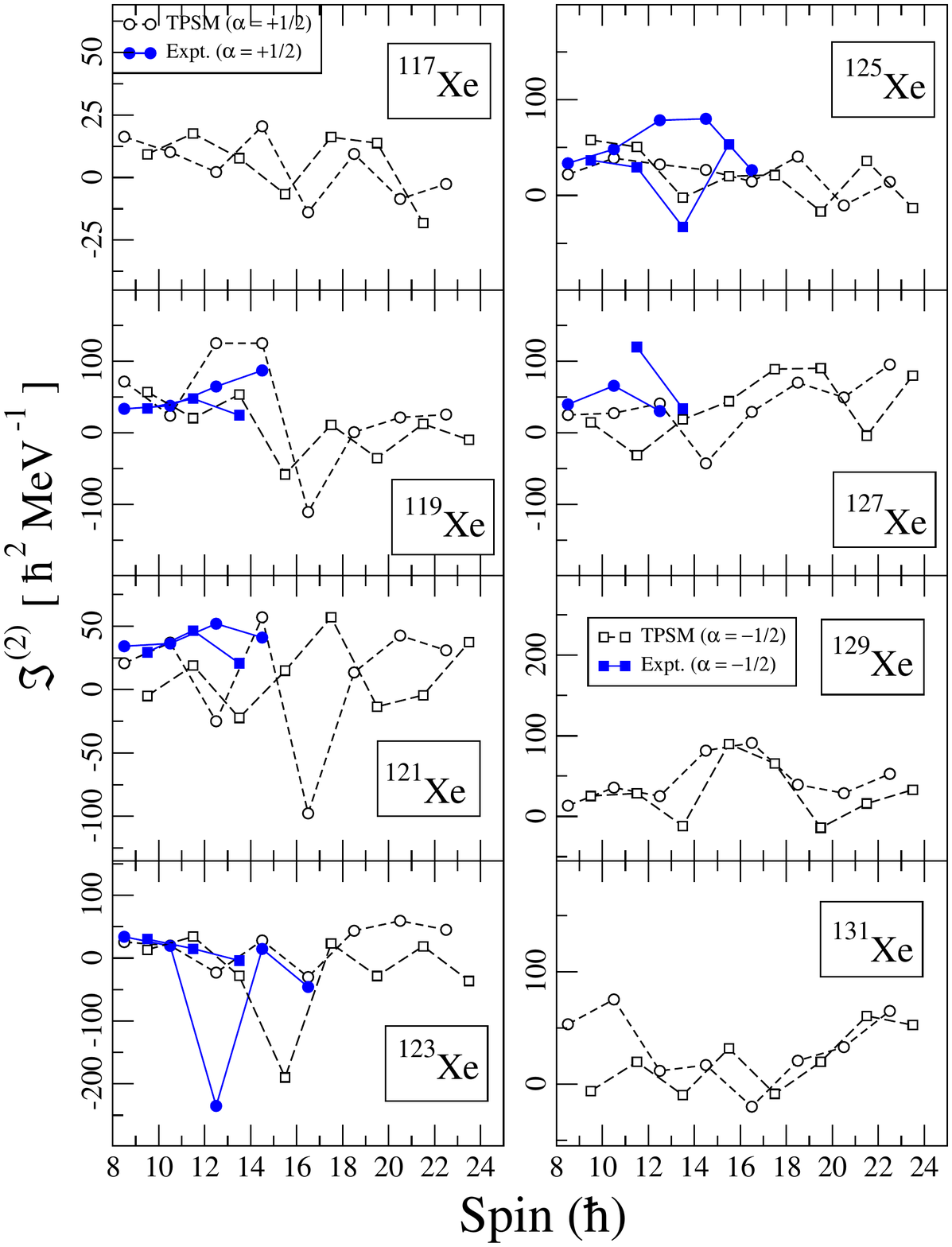}}
\caption{
Comparison between experimental and calculated dynamic moment of inertia   
of the yrare-band for$^{117-131}$Xe isotopes.}\label{ali2yrare}
\end{figure}
\begin{figure}[htb]
 \centerline{\includegraphics[trim=0cm 0cm 0cm
0cm,width=0.53\textwidth,clip]{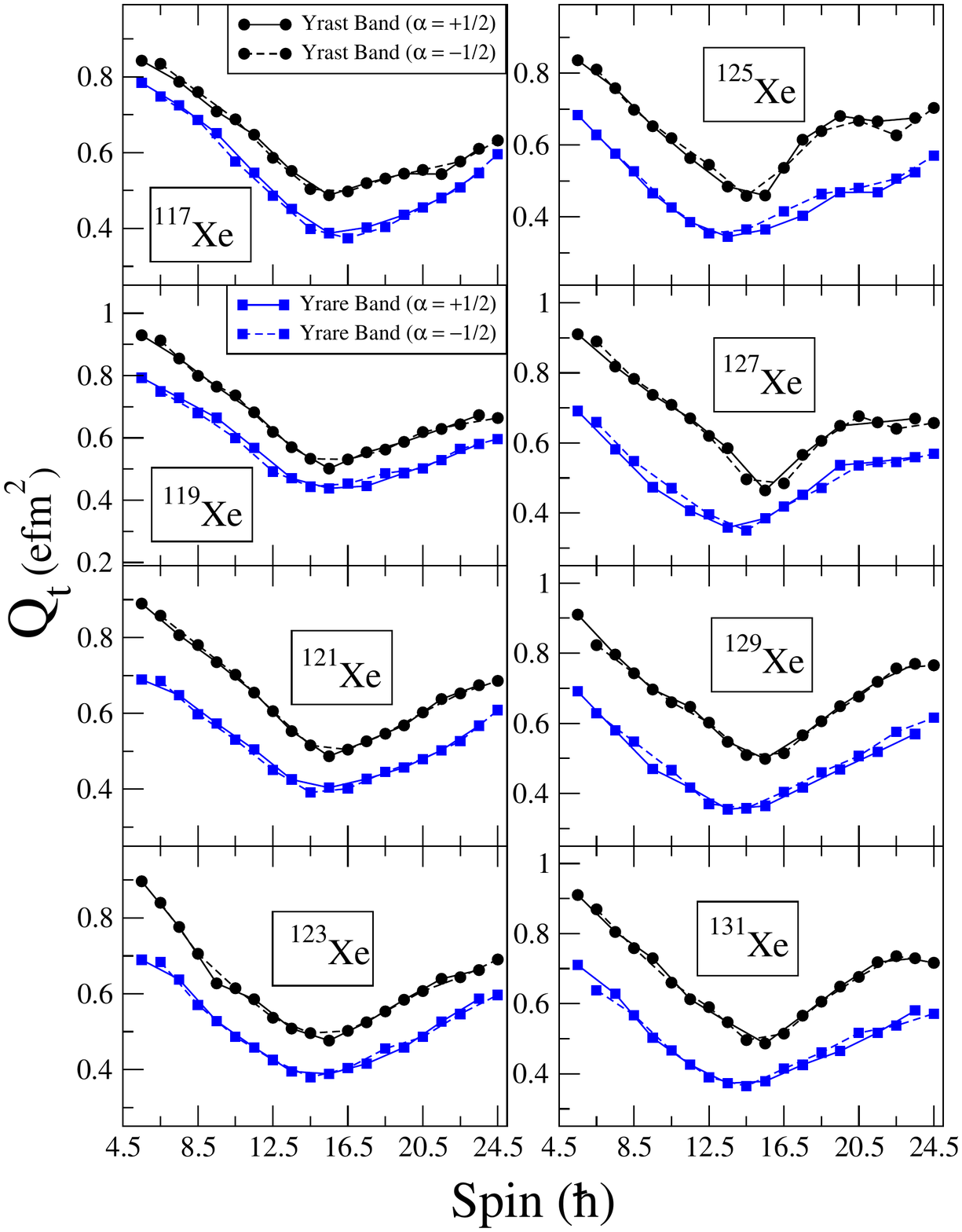}}
\caption{
Calculated transition quadrupole moment, $Q_t (efm^2)$, 
for yrast and yrare bands in $^{117-131}$Xe isotopes.} 
\label{be21}
\end{figure}

In recent years, the triaxial projected shell model (TPSM) approach
has been demonstrated to provide a unified description of the
high-spin band structures of rotational and transitional nuclei with 
a remarkable accuracy \cite{SH99,JS16,JS21}. The advantage of this model is that basis space
is composed of  angular-momentum projected multi-quasiparticle states
which allows to investigate the band structures up to
quite high spin. The basic tenet of the TPSM approach is quite
similar to that of spherical shell model approach with the only
exception that deformed basis are employed as compared to the
spherical states. The deformed states form the optimum basis to 
investigate the properties of deformed nuclei. 

In the earlier version of the TPSM approach, the basis space was quite
limited, and it was not possible to investigate the high-spin
states \cite{GH08,JG09,SB10,JG11,GJ12,JG12,JM17}. For odd-neutron systems, the
basis space was one-neutron and one-neutron coupled to two-proton states. In the present work, we have 
expanded the basis space to include three-neutron and three-neutron 
coupled to two-proton configurations.  This extension makes it feasible to
investigate the odd-neutron systems up to quite high angular
momentum and as a first major application of this development, the
high-spin band structures for odd-neutron $^{117-131}$Xe have
been investigated in the present work. The present work is an ongoing effort of our group
to extend the TPSM approach and in our recent publications, we have expanded the basis space of 
even-even \cite{SJ18}, odd-odd \cite{sjepja}  and odd-proton \cite{JS21b} systems. We would like 
to mention that in a parallel effort, the axial version of the TPSM approach
has also been generalized to multi-quasiparticle states \cite{PhysRevC.90.011303,PhysRevC.93.034322,PhysRevC.97.044302}.

The band structures in odd-mass Xe-isotopes have been extensively 
investigated using the state-of-the-art experimental techniques \cite{123Xe-2020,125Xe,127Xe-2020}. 
In some of these studies, the high-spin states have been populated up to spin, $ I \approx 55\hbar $
and also several side bands have been identified \cite{123Xe,125Xe}. The major problem is
how to characterise the intrinsic configurations of the observed band structures
as single-particle and collective degrees of freedom are
interwoven in odd-mass systems. 
The extended TPSM approach developed
in the present work provides a unified description of the
single-particle and collective modes. It will be established in the
present study that some excited bands observed in these nuclei are 
$\gamma$ bands built on quasiparticle states. The observation of the signature
inversion in the $\gamma$ bands in some of the studied nuclei 
will also be addressed in the present work. 

Further, it has been observed in the odd-neutron Xe-isotopes that the 
character of the band crossing changes from proton to neutron with increasing
mass number \cite{123Xe-2020,125Xe,127Xe-2020,lk18}. As in the extended model space, both proton
and neutron quasiparticle configurations are included, it is possible
to investigate the interplay between neutron and proton
alignments. This interplay between different quasiparticle
configurations shall be studied in detail in the present work. 
The manuscript is organised in the following 
manner. In the next section, the extended TPSM approach is briefly
presented. In section III, TPSM results obtained for odd-mass Xe-isotopes
are compared with the experimental data, wherever
available. Finally, the results obtained in the present work are summarised 
and concluded in section IV.

\section{Extended Triaxial Projected Shell Model Approach}

To provide a microscopic description of collective and multi-quasiparticle excitations in
odd-neutron systems, TPSM approach has been extended by including 
three- and five-quasiparticle basis states.
Odd-neutron systems have been studied earlier using the TPSM approach with the restricted model
space of one-neutron and one-neutron coupled to
two-proton quasiparticle states. However, in order to investigate the
high-spin data, which is now available for many odd-neutron systems,
neutron aligning configurations are needed, in addition to
the proton aligning states. In the present work, the extended basis space has
been implemented, and the complete basis space in the generalised approach is given by $:$
\begin{equation}
\begin{array}{r}
~~\hat P^I_{MK}~ a^\dagger_{\nu_1} \ack\Phi\ket;\\
~~\hat P^I_{MK}~a^\dagger_{\nu_1}a^\dagger_{\pi_2}a^\dagger_{\pi_3}  \ack\Phi\ket;\\
~~\hat P^I_{MK}~a^\dagger_{\nu_1}a^\dagger_{\nu_2}a^\dagger_{\nu_3}  \ack\Phi\ket;\\
~~\hat P^I_{MK}~a^\dagger_{\nu_1}a^\dagger_{\nu_2} a^\dagger_{\nu_3}a^\dagger_{\pi_1} a^\dagger_{\pi_2} \ack\Phi\ket,
\label{intrinsic}
\end{array}
\end{equation}
where $\ack\Phi\ket$ is the triaxially deformed quasiparticle vacuum state, and
$P^I_{MK}$ is the three-dimensional
angular-momentum-projection operator given by \cite{RS80} $:$ 
\begin{equation}
\hat P ^{I}_{MK}= \frac{2I+1}{8\pi^2}\int d\Omega\, D^{I}_{MK}
(\Omega)\,\hat R(\Omega),
\label{Anproj}
\end{equation}
with the rotation operator 
\begin{equation}
\hat R(\Omega)= e^{-i\alpha \hat J_z}e^{-i\beta \hat J_y}
e^{-i\gamma \hat J_z}.\label{rotop}
\end{equation}
Here, $''\Omega''$ represents the set of Euler angles 
($\alpha, \gamma = [0,2\pi],\, \beta= [0, \pi]$) and  
$\hat{J}^{,}s$ are the angular-momentum operators.

The constructed projected basis of Eq.~(\ref{intrinsic})
is then used to diagonalize the shell model Hamiltonian. In this sense, the present approach is analogous
to the standard shell model (SM) approach with the difference that projected deformed basis is employed as compared
to the spherical basis in the SM approach. In the present work, the shell model Hamiltonian consists of a sum of
quadrupole-quadrupole, monopole pairing and quadrupole pairing interaction terms. These terms describe the
principle components of the nuclear potential \cite{BarangerKumar, dufour}. The Hamiltonian is given by $:$
\begin{eqnarray}
\hat H =  \hat H_0 -   {1 \over 2} \chi \sum_\mu \hat Q^\dagger_\mu
\hat Q^{}_\mu - G_M \hat P^\dagger \hat P - G_Q \sum_\mu \hat
P^\dagger_\mu\hat P^{}_\mu . \label{hamham}
\end{eqnarray}
In the above equation, $\hat H_0$ is the spherical single-particle
part of the  Nilsson potential \cite{Ni69}. 
As a consequence of the self-consistent HFB condition, the QQ-force strength, $\chi$, in Eq. (\ref{hamham}) is related to
the quadrupole deformation $\epsilon$. The monopole pairing strength $G_M$ (in MeV)
is of the standard form
\begin{eqnarray}
G_M = {{G_1 \mp G_2{{N-Z}\over A}}\over A}, 
 \label{pairing}
\end{eqnarray}
where the minus sign applies to neutrons, and plus sign applies to protons. In the present
work, $G_1$ and $G_2$ are fixed such that the 
calculated gap parameters approximately reproduce the experimental odd-even mass differences in the
mass region under investigation. 
The single-particle space employed 
in the present calculation is three major oscillator shells (N$=3,4,5$) for both neutrons and protons.
The  strength $G_Q$ for the quadrupole pairing force is fixed as 0.16 times the $G_M$ strength.
These interaction strengths are consistent with those used earlier in the TPSM calculations \cite{JG12,KY95,sh10}.

To diagonalize the shell
model Hamiltonian,  Eq.~(\ref{hamham})  in the angular-momentum projected basis, 
the Hill-Wheeler approach is followed \cite{KY95}. The generalized eigen-value
equation is given by 
\begin{equation}
 \sum_{\kappa^{'}K^{'}}\{\mathcal{H}_{\kappa K \kappa^{'}K^{'}}^{I}-E\mathcal{N}_{\kappa K 
\kappa^{'}K^{'}}^{I}\}f^{\sigma I}_{\kappa'K'}=0, \label{a15}
\end{equation}
 where the Hamiltonian and norm kernels are given by
 \begin{eqnarray*}
 && \mathcal{H}_{\kappa K \kappa^{'}K^{'}}^{I} = \langle \Phi_{\kappa}|\hat H\hat 
P^{I}_{KK^{'}}|\Phi_{\kappa^{'}}\rangle ,\\
&&\mathcal{N}_{\kappa K \kappa^{'}K^{'}}^{I}= \langle \Phi_{\kappa}|\hat P^{I}_{KK^{'}}|\Phi_{\kappa^{'}}\rangle .
 \end{eqnarray*}
 The Hill-Wheeler wave function is given by
\begin{equation}
\psi^{\sigma}_{IM} = \sum_{\kappa,K}~f^{\sigma I}_{\kappa K}~\hat P^{I}_{MK}| 
~ \Phi_{\kappa} \ket.
\label{Anprojaa}
 \end{equation}
where $f^{\sigma I}_{\kappa K}$ are the variational coefficients and index  $^{``}\kappa^{``}$ designates the basis states of Eq. (\ref{intrinsic}).
The wave-function is then used to evaluate the electromagnetic 
transition probabilities. 
The reduced electric transition probabilities $B(EL)$ from an initial state 
$( \sigma_i , I_i) $ to a final state $(\sigma_f, I_f)$ are given by \cite{JS21}
\begin{equation}
 B(EL,I_i \rightarrow I_f) = {\frac {1} {2 I_i + 1}} 
| \bra \ \psi^{\sigma_f I_f}|| \hat Q_L || \psi^{\sigma_i I_i} \ket |^2 ,
  \end{equation}
and the reduced matrix element can be expressed as
\begin{eqnarray*}
& \bra &\ \psi^{\sigma_f I_f}|| \hat Q_L || \psi^{\sigma_i I_i} \ket
\nonumber \\ 
&=&\sum_{\kappa_i , \kappa_f, K_i,K_f} {f_{ \kappa_i K_{i}}^{\sigma_i I_i}}~ {f_{ \kappa_f K_{f}}^{\sigma_f I_f}}
 \sum_{M_i , M_f , M} (-)^{I_f - M_f}  \nonumber \\&\times&
\left(
\begin{array}{ccc}
I_f & L & I_i \\
-M_f & M &M_i 
\end{array} \right) 
\nonumber \\
 & & \times \bra \Phi | {\hat{P}^{I_f}}_{K_f M_f} \hat Q_{LM}
\hat{P}^{I_i}_{K_i M_i} | \Phi \ket 
\nonumber \\
 &=& 2 \sum_{\kappa_i , \kappa_f,K_i,K_f} {f_{ \kappa_i K_{i}}^{\sigma_i I_i}}~ {f_{ \kappa_f K_{f}}^{\sigma_f I_f}}
\nonumber \\
 & & \times \sum_{M^\prime,M^{\prime\prime}} (-)^{I_f-K_f} (2 I_f + 1)^{-1}
\left( 
\begin{array}{ccc}
I_f & L & I_i \\
-K_{f} & M^\prime & M^{\prime\prime}
\end{array} \right)\\
 & & \times \int d\Omega \,D^{I_i}_{M''K_{i}}(\Omega)
\bra \Phi_{\kappa_f}|\hat { O}_{LM'}\hat R(\Omega)|\Phi_{\kappa_i}\ket. 
\end{eqnarray*}

In the present work, we have evaluated the transition quadrupole moment, $ Q_t (I) $, which is related to $B(E2)$ transition 
probability through  
\begin{equation}
Q_t(I) = \sqrt{\frac{16 \pi}{5}} \frac{ \sqrt{B(E2,I\rightarrow I-2)}} 
{\bra I, K, 2, 0 | I-2, K\ket } . 
\end{equation}
In the numerical calculations, 
we have used the standard effective charges of 1.5e for protons 
and 0.5e for neutrons \cite{KY95,Sun94}.

\section{Results and Discussion}
 \begin{table}[!b]
\caption{Axial and triaxial quadrupole deformation parameters
$\epsilon$ and $\epsilon'$  employed in the TPSM calculation.}
\begin{tabular}{|ccccccccc|}
\hline                   & $^{117}$Xe &  $^{119}$Xe    &  $^{121}$Xe  &$^{123}$Xe  & $^{125}$Xe& $^{127}$Xe  & $^{129}$Xe & $^{131}$Xe   \\
\hline   $\epsilon$      &0.234      &  0.227       &  0.209      & 0.220    & 0.180    &  0.150    &  0.150   &   0.160    \\
\hline $\epsilon'$       &0.110      &  0.100       &  0.100      & 0.105     & 0.090    &  0.100    &  0.095   &  0.090       \\
\hline $\gamma $          & 25      &   24        &   24        & 24         & 27        & 33        & 32       &   29        \\
\hline
\end{tabular}
\label{tab1}
\end{table}
TPSM calculations have been performed for eight odd-mass $^{117-131}$Xe 
isotopes using the axial and non-axial deformations listed in 
Table 1. These deformation values have been adopted from the earlier 
studies performed for these nuclei \cite{117xeb,119xeb,Tim_r_1995,123Xe-2020,125Xe,127Xe-2020,129xed,131xed}. The angular
momentum projected energies for the configurations in the vicinity
of the Fermi surface are depicted in Figs. \ref{bd1}, \ref{bd2a} and
\ref{bd2b} for the studied isotopes. These plots, referred to as the band diagrams, provide important
information on the intrinsic structures of the observed band
structures, which in turn sheds light on the nature of band crossings. For
$^{117}$Xe, the projected energies from the lowest $I=7/2$ to 47/2 are depicted in
Fig.~\ref{bd1}. These diagrams are similar for all other studied
Xe-isotopes and only the interesting band crossing portion of the
diagrams are shown in Figs.~\ref{bd2a} and \ref{bd2b} since the crossing
features vary from isotope to isotope. 

The ground-state band for $^{117}$Xe is the projected band from the
one-quasineutron configuration having $K=3/2$ and intrinsic energy of 1.25 MeV. 
The projection from this
triaxial intrinsic state also leads to several other band structures
with $K=7/2$ and 11/2, which are so called $\gamma$ and $\gamma\gamma$ 
bands built on the $K=3/2$ state. The band heads of these bands are located at excitation
energies of $1.01$ MeV and $2.10$ MeV, respectively. In recent years, several
excited bands have been observed in Xe-isotopes, and some of these
bands have been conjectured to be the $\gamma$ bands. It is one of the
objectives of the present work to investigate these
structures as candidate $\gamma$ bands in odd-mass Xe-isotopes.

It is observed from Fig.~\ref{bd1} that three-quasiparticle band 
with $K=3/2$ configuration crosses the
ground-state band at $I=33/2$. This crossing of the configuration having 
one-neutron coupled to two-protons will correspond to the first
band crossing observed for this system. It is quite interesting
to note that $K=7/2$, which is the $\gamma$ band based on the three
-quasiparticle state also crosses the normal $\gamma$ band built on 
the ground-state band. This is expected since $\gamma$ bands are projected from the same
intrinsic state as that of the parent state, but with a different value of "$K$" quantum
number. Further, this band also crosses the ground-state band at a slightly
higher angular momentum and it is, therefore, predicted that at high-spin
two parallel band structures should be observed, one corresponding
to the normal one-neutron plus two-proton configuration and the other to the
$\gamma$ band based on it. 

The segments of the band diagrams for other Xe-isotopes that 
include only the band crossing regions are displayed in Figs.~\ref{bd2a}
and \ref{bd2b}. 
For $^{119}$Xe, the band crossing noted at $I=33/2$ is due to alignment
of neutrons rather than that of protons as for $^{117}$Xe. 
For $^{121,123,125,127}$Xe isotopes, the first crossing is again due
to neutrons, however, for $^{129,131}$Xe isotopes the alignment is 
due to protons as for the lighter isotope of $^{117}$Xe. In most of the 
isotopes, it is noted that three-quasiparticle state having $K=3/2$ and the
$\gamma$ band built on it, almost simultaneously cross the ground-state
band. This has been also found in several even-even systems in this region \cite{ee1,ee2,ee3}. 

High-spin states in odd-neutron Xe-isotopes have been investigated by
various  experimental groups \cite{117xeb,119xeb,123Xe-2020,125Xe,127Xe-2020,129xed,131xed}, and
in some nuclei side bands apart
from the yrast states have been populated up to quite high angular
momentum. In Figs. ~\ref{expe1} to \ref{expe4}, the calculated band structures obtained
after diagonalization of the shell model Hamiltonian are compared with
the observed energies. TPSM band structures are plotted for the
yrast, yrare and also for the three-quasiparticle excited band as in
some of the studied nuclei excited bands have been observed. The
purpose here is to elucidate the intrinsic structures, where these
bands have already been identified. For other nuclei, the predicted band structures 
will provide some guidance for future experimental investigations. 

For $^{117}$Xe, the yrast band is known up to $I=43/2$ for the favoured
signature ($\alpha=1/2$) and up to $I=33/2$ for the unfavoured
signature ($\alpha=-1/2$). The calculated TPSM energies, shown in
Fig.~~\ref{expe1}, are noted to be in good agreement with the known energies.
The deviation between experimental and the calculated energy for 
the highest observed spin state is about 0.15 MeV. 
In Fig.~\ref{expe1}, the TPSM energies are also given for the yrare and the second 
excited band.  In some Xe-isotopes, excited band
structures have been observed and we hope that in future experimental
studies, excited structures will also be identified for
$^{117}$Xe. Fig.~\ref{expe1} also compares the known experimental bands with
the TPSM calculated energies for $^{119}$Xe. For this system, apart from the yrast
band  that is observed up to $I=43/2$, yrare band is also known up to
$I=29/2$. It is evident from the figure that TPSM calculations reproduce the
experimental energies fairly well. The unfavoured branch of the
$\gamma$ band is lower in energy as compared to the favoured branch for low-spin, and
then at higher spin the situation is reversed. The signature splitting of the $\gamma$ bands will be discussed in
detail later.

The calculated band structures for $^{121}$Xe and $^{123}$Xe are
compared with the known energies in Fig.~\ref{expe2}. Yrare band up to
$I=29/2$ is observed in $^{121}$Xe, and in $^{123}$Xe apart from the 
yrare band, one more excited band is known. The band structures
for the isotopes of $^{125}$Xe, and $^{127}$Xe are compared in
Fig. ~\ref{expe3}, and it is noted that agreement between the TPSM
and the known experimental energies is quite reasonable. For
both the isotopes, apart from the yrare band, one more excited 
band is known with the band head at $I=27/2$. The results for 
$^{129}$Xe and $^{131}$Xe are compared in Fig.~\ref{expe4}, and again the 
TPSM energies are in good agreement with the known
energies. For both these isotopes, excited bands have been observed,
which are quite high in energy and appear to be based on three
-quasiparticle configuration. In the following, we shall examine the 
intrinsic structures of the bands,
presented in Figs.~\ref{expe1} to \ref{expe4}, through 
the analysis of the wave functions.

The wave function amplitudes of the yrast, yrare and the second excited 
bands are displayed in Figs.~\ref{wf2}, \ref{wf3} and \ref{wf4}, respectively for
the eight studied isotopes. The yrast band at low-spin has the
dominant contribution from the projected one-quasineutron 
configuration with $K=3/2$. At high-spin, three-quasiparticle state
which crosses the ground-state band becomes dominant. For $^{117}$Xe,
the one-neutron coupled to two-proton 
configuration crosses and, therefore, the first
band crossing is due to the alignment of protons. For the isotopes
from $^{119}$Xe to $^{127}$Xe, it is three-quasineutron configuration that crosses the ground-state band and the first
crossing for these isotopes is due to alignment of neutrons. For the
two heavier isotopes of $^{129}$Xe and $^{131}$Xe, the crossing is
due to protons as for $^{117}$Xe. It is also noted from Fig.~\ref{wf2}
that five-quasiparticle state becomes important at high-spin, in
particular, for $^{117}$Xe. 

The amplitudes for the yrare band, Fig.~\ref{wf3}, which is a $\gamma$ band in the
low-spin region also depicts a
crossing phenomena similar to that of the yrast band. 
For the yrast band, the band
crossing is due to the alignment of three-quasiparticle and in
the case of the yrare band, it is the $\gamma$ band built on this
parent three-quasiparticle state. What is interesting is that
angular-momentum at the crossing point is similar for the yrast and the
$\gamma$ band.
Therefore, $\gamma$ band tracks the yrast band for the studied
odd-neutron Xe-isotopes as has been observed
for the even-even $^{156}$Dy nucleus \cite{156Dy}.

In Fig.~\ref{wf4}, the wave function amplitudes for the second excited band are shown
for the eight Xe-isotopes. These bands have been observed for some of the
studied isotopes. In general, it is noted from the figure that the wave function 
of the excited
band have mixed intrinsic compositions due to high density of states at higher 
excitation energies. For $^{117}$Xe, $^{119}$Xe, $^{121}$Xe, $^{123}$Xe, $^{125}$Xe and $^{127}$Xe, low-spin states have
dominant three-neutron configuration, and at high-spin five-quasiparticle state becomes
important. One-neutron coupled to two-proton configuration is dominant for $^{129}$Xe
and $^{131}$Xe isotopes.

It is expected that in future experimental studies many three-quasiparticle band structures will be identified at
high-spin. To provide some guidance to these studies, we have investigated the angular momentum dependence of the
lowest few three-quasiparticle bands that become favoured at high-spin. The band heads of these bands are depicted
in Fig.~\ref{bhe1} with reference to the yrast state for each angular momentum. These  energies have been
calculated after diagonalization of the shell model Hamiltonian and 
have mixing from various quasiparticle states. The dominant
component for each band head is indicated in the legend of Fig.~\ref{bhe1}. 
It is quite evident from the
figure that for low-$I$, these states are quite high in excitation
energy, but become favoured in the high-spin region. In particular, it
is noted that $\gamma$ band built on the three-quasiparticle states
come close to the yrast line at high-spin.

The observation of 
anomalous signature splitting of negative parity bands in odd-\textit{A} Xe isotopes has attracted a
considerable attention in recent years \cite{125Xe_neg-parity,119xeb}. There are four negative parity
rotational bands, \textit{viz.} favoured and unfavoured signature partners ($ \alpha = \pm 1/2 $) of the yrast and
yrare bands, based on the $ \nu h_{11/2} $ orbital that are reported systematically in odd-\textit{A} Xe isotopes. The
yrast band was reported in $ ^{117-131} $Xe with large signature splitting. The origin of this band is explained in
terms of the coupling of a quasineutron in the $ h_{11/2} $ orbital to the ground state configuration of the
core (\textit{i.e.}, $ \nu h_{11/2} \otimes 0^{+}_{1}| $\textsubscript{\textsuperscript{even}Xe}). However, the
observed large signature splitting in this band is quite unexpected for a band associated with high-$ \Omega $
quasiparticles ($ \Omega \geq \frac{3}{2} $). Theoretical calculations predict  that the signature splitting of
yrast negative parity bands is very sensitive to the $ \gamma $-deformation \cite{TRPM_Xe-Ba}. For instance,
the observed $ S(I) $ in the case of $^{125}$Xe ($ [523]\frac{7}{2} $) is reproduced well with
$ \gamma \approx 24^{\circ} $ \cite{125Xe_TPRM}. However, the $ S(I) $ is found to be normal
in the case of positive parity bands, in spite of having similar $ \gamma $-deformation
\cite{127Xe_pos-parity}. In contrast to the yrast bands, the yrare bands, which are thought to
originate from the coupling of an $ h_{11/2} $ neutron with the $ \gamma $-vibration of the
core (\textit{i.e.}, $ \nu h_{11/2} \otimes 2^{+}_{2}| $\textsubscript{\textsuperscript{even}Xe}),
in $ ^{119-125} $Xe were reported with a low, fairly constant and inverted signature
splitting \cite{125Xe_neg-parity}. But, the $ S(I) $ of the quasi-$ \gamma $-bands in
\textsuperscript{even}Xe isotopes are not inverted and also varies with the
mass number \cite{126Xe_2020}. Therefore, such a simple coupling scheme is
not adequate enough to explain the origin of the yrare bands. 

To shed light on the observation of signature inversion in some odd-mass Xe isotopes, we have evaluated the signature
splitting, $S(I)$, of the yrast and the $\gamma$ bands using the TPSM energies. The calculated signature splitting and the
corresponding experimental values for the two bands are displayed in Figs.~\ref{sp1y} and \ref{sp2y}. It is evident from
Fig.~\ref{sp1y} that the calculations reproduce the experimental signature
splitting for the yrast band quite well, and is a validation that $\gamma$ deformation 
values employed in the TPSM model are reasonable. 
For the isotopes of  
$^{119}$Xe and $^{125}$Xe, the yrare band is known up to high-spin and it is observed that
favoured signature lies higher in energy than the unfavoured
one for the low-spin states, and then around $I=12$ signature inversion is
noted. It is evident from Fig.~\ref{sp2y} that signature inversion is
well reproduced by the TPSM calculations, and is readily understood as
due to the crossing of the three-quasiparticle band with the
$\gamma$ band as is seen from the
band diagrams, Figs.~\ref{bd1}, \ref{bd2a} and \ref{bd2b}. The two crossing configurations have opposite
signature phase and gives rise to signature inversion. For $^{119}$Xe, the TPSM calculated $S(I)$ depicts another
signature inversion at about $I=17$ and is due to the interaction with other quasiparticle configurations.

It is noted that the three-quasiparticle band crosses the $\gamma$ bands in all the 
Xe-isotopes and it is, therefore, expected that for all the studied nuclei, $\gamma$ band
will depict a change in the signature phase. For $^{121}$Xe and $^{123}$Xe, the observed favoured signature
again lies at a higher energy than the unfavoured one, but no inversion is noted. The inversion is seen at a higher angular momentum
in the TPSM calculated $S(I)$. For $^{127}$Xe, the inversion is observed at a slightly lower angular momentum and is
again well reproduced by the TPSM calculations. For the three isotopes of $^{117}$Xe, $^{129}$Xe and $^{131}$Xe, yrare bands
have not been observed, but the theoretical calculations predict a similar behaviour of $S(I)$ as for other isotopes. The
yrast band $S(I)$, shown in Fig.~\ref{sp1y}, does not depict any signature inversion
at the crossing point as the phase of the signature splitting 
of both the bands is same, although some modification in the signature splitting is noted after the band crossing since
the two bands have different $S(I)$.

We shall now turn to the discussion of the band crossing features for the studied
Xe-isotopes. It has been demonstrated in several studies that nature of the first band crossing
changes with the shell filling \cite{lk18, cexe, ysunpr}. In order to investigate the detailed features of 
crossing phenomena, we have
calculated the quantities $:$ aligned angular momentum ($i_x$) and
dynamic moment of inertia ($J^{(2)}$). These quantities are displayed
in Figs.~\ref{ali1}, \ref{ali1yrare},\ref{ali2} and \ref{ali2yrare} for the yrast and the yrare bands.
In the caption of Fig.~\ref{ali1} and Fig.~\ref{ali2}, the expressions and the
parameters used to evaluate $i_x$ and $J^{(2)}$ are provided. For the yrast bands of $^{119}$Xe,
$^{123}$Xe, $^{129}$Xe and $^{131}$Xe, $i_x$ in Fig. ~\ref{ali1} depict back-bends,
indicating that the crossing between the bands is weak. For the other
isotopes, $i_x$ depict up-bends, which suggests that the interaction between the
two bands is large. For $^{125}$Xe, back-bend is noted at a higher rotational frequency.
The TPSM calculated $i_x$ for the yrare band, displayed
in Fig.~\ref{ali1yrare}, depict back-bends in all the cases. The experimental values are known only
in the low-spin regime and TPSM calculations reasonably reproduce these values. The dynamic moment of
inertia values, compared in Figs.~\ref{ali2} and \ref{ali2yrare} for the yrast and the yrare bands, show a reasonable agreement 
between the TPSM calculated numbers and those deduced from the 
experimental data. TPSM calculated $J^{(2)}$ for the yrare band depicts large structural
changes for all the studied isotopes as this band interacts with many other bands. 

We have also studied the transition quadruple moment, $Q_t$, along the yrast and yrare bands since the
quasiparticle alignments are expected to give rise to deformation changes. The calculated $Q_t$ for the
two bands are depicted in Fig.~\ref{be21} and it is noted that for all the isotopes,  $Q_t$ drops in the
band crossing region. This drop is expected since in this region, the wave function is a
mixture of ground and the aligning configurations. Further, it is noted that yrast and
yrare bands have similar behaviour, and is easily understood since both the bands originate
from the same intrinsic configuration and also have similar band crossing features.
The difference in the magnitudes of $Q_t$ for the two bands can be mainly attributed to the different ``$K$'' composition in the
two bands - the yrast band is dominated by $K=3/2$ whereas the yrare band has the $K=7/2$ predominant component.

\section{Summary and Conclusions}

In the present work, the triaxial projected shell model approach has been extended
to include three-neutron and five-quasiparticle configurations for odd-neutron systems. 
This generalisation has made it feasible to investigate the intrinsic structures of the observed excited bands
in odd-neutron systems. For odd-mass Xe-isotopes, several excited band structures have
been observed, and the configurations of these bands have been discussed. As the protons and neutrons
occupy the same configuration space, the interplay between them plays a crucial role to determine the
structures of the observed bands. In the earlier version of the TPSM approach, the basis space for odd-neutron
systems was comprised of one-neutron and one-neutron coupled to two-proton state, and it was not
possible to study the interplay between neutron and proton aligned configuration. It is known from the CSM
analysis \cite{lk18} that the nature of the band crossing changes from proton to neutron with the $1h_{11/2}$ shell
filling.  It has been elucidated using the extended model space that band crossing for $^{117}$Xe is due to the
alignment of protons, and for $^{119}$Xe to $^{127}$Xe it is due to the alignment of neutrons. For the two isotopes
of $^{129}$Xe and $^{131}$Xe, it is again due to the alignment of protons. 

Further, it has been demonstrated that the excited bands observed in some odd-neutron Xe-isotopes
are actually $\gamma$ bands based on three-quasiparticle configurations. It has been discussed in several
TPSM studies that $\gamma$ bands are built on each quasiparticle states as for the ground-state band. For even-even
systems, several s-bands have been identified, and it has been observed in many cases that $g$-factors
have similar values for the
band head $10^+$ states \cite{ee1}. The similar nature of the $g$-factors was quite surprising as normally one expects
different $g$-factors for proton and neutron aligning configurations. In this region, both neutrons and
protons tend to align almost
simultaneously as the two Fermi surfaces are in close vicinity. It is then expected that one s-band should
have neutron
character, and the other to have the proton structure. The observation of  similar $g$-factors was
puzzling as they should
be different, corresponding to protons and neutrons. It was clarified using the TPSM approach that the two observed
s-bands are actually two-particle aligned configuration and the $\gamma$ band based on
this aligned state \cite{sj17}. Since
the two bands have same intrinsic structure and it is expected that $g$-factors of the two s-bands
should have
similar values. It was shown for $^{134}$Ce that the two bands have negative $g$-factors i.e., neutron character,
and for $^{136}$Nd it was demonstrated that they
have proton character as the two lowest aligned bands have positive $g$-factor \cite{sj17}. 

For odd-mass systems, we also expect a similar band crossing phenomenon in this region. It
has been shown in the present work that three-quasiparticle band crosses the one-quasiparticle 
ground-state band and leads to the standard band crossing phenomenon observed along the yrast line.
The interesting inference from the  present work is that normal $\gamma$ band, which is the first excited band,
is also crossed by $\gamma$ band based on the three-quasiparticle configuration that crosses the ground-state
band. Therefore, $\gamma$ band tracks the ground-state band with analogous band crossing occurring in the two
bands \cite{156Dy}. It has been further observed that some $\gamma$ bands in Xe-isotopes depict signature
inversion at high-spin, and it is now quite evident from the present investigation that this inversion is directly
related to the occurrence of the band crossing along the $\gamma$ band. The three-quasiparticle band that crosses
the ground state band at high spin, first crosses the $\gamma$ band at a lower spin and give rise to
signature inversion as the two bands have opposite phase of the signature splitting.

We have also provided excitation energies of the three-quasiparticle configurations, which become favoured
at high-spin. Some of these band structures have already been observed in a few isotopes, and we hope that in future 
experimental work many more excited bands will be identified. It will be interesting to measure the $g$-factors of
these excited bands at high-spin as some of these bands are $\gamma$ bands based on three-quasiparticle
configurations and will have similar $g$-factor as that of the parent band. Further, the transition quadrupole moment
has been studied, and it has been demonstrated that both yrast and yrare bands have similar behaviour as a
function of spin.

\section{ACKNOWLEDGEMENT}
The authors  would like to acknowledge Science and Engineering Research Board (SERB), Department of Science and
Technology (Govt. of India) for providing financial assistance under the 
Project No.CRG/2019/004960 to carry out a part of the present research work. 

\end{document}